\pdfoutput=1

\documentclass[preprint2,longabstract]{aastex}

\usepackage[english]{babel} 

\usepackage{url} 
\usepackage{subfigure}
\usepackage{longtable}
\usepackage{lscape}
\usepackage{multirow}
\usepackage{color}
\usepackage{textcomp}
\usepackage{natbib}
\usepackage{epsfig}
\usepackage[colorlinks,linkcolor=red,anchorcolor=green,citecolor=blue]{hyperref}

\usepackage{graphicx}
\usepackage{txfonts}
\usepackage{epstopdf}
\usepackage{float}
\usepackage[absolute]{textpos}

\newcommand{\CCH}{C$_2$H}
\newcommand{\NNHp}{N$_2$H$^+$}
\DeclareRobustCommand{\ion}[2]{\textup{#1\,\textsc{\lowercase{#2}}}}

\begin{document}

\title{\CCH~$N=1-0$ and \NNHp~$J=1-0$ observations of Planck Galactic cold clumps}

\author{X.-C. Liu \altaffilmark{1,2},
        Y. Wu\altaffilmark{1,2},
C. Zhang\altaffilmark{3,1,2},
T. Liu\altaffilmark{4},
J. Yuan\altaffilmark{5},
S.-L. Qin\altaffilmark{3},
B.-G. Ju\altaffilmark{6,7},
L.-X. Li\altaffilmark{2}
}
   \altaffiltext{*}{[1501110219;ywu]@pku.edu.cn}
   \altaffiltext{1}{Department of Astronomy, Peking University, China}
   \altaffiltext{2}{KIAA, Peking University, 100871 Beijing, China}
   \altaffiltext{3}{Department of Astronomy, Yunnan University, China}
   \altaffiltext{4}{Korea Astronomy and Space Science Institute, Korea}
   \altaffiltext{5}{NAOC, Beijing 100101, China}
   \altaffiltext{6}{PMO, Qinghai Station, 817000, Delingha, China}
   \altaffiltext{7}{Key Laboratory for Radio Astronomy, CAS}

\begin{abstract}
A survey of  C$_2$H $N=1-0$  and   N$_2$H$^+$ $J=1-0$
toward Planck Galactic cold clumps (PGCCs) was performed
using the Purple Mountain Observatory's 13.7 m telescope.
C$_2$H and  N$_2$H$^+$ were chosen to study the chemical evolutionary states of PGCCs.
Among 121 observed molecular cores associated with PGCCs,
71 and 58  are detected with C$_2$H $N=1-0$ and N$_2$H$^+$ $J=1-0$, respectively.
The detected lines of most sources can be fitted with a single  component with compatible
V$_{\mathrm{LSR}}$ and line widths, which confirms that these PGCC cores are very cold
(with gas temperatures 9--21 K)
and quiescent while still dominanted by turbulence.
The ratio between the column densities of C$_2$H and N$_2$H$^+$ ($N$(C$_2$H)/$N$(N$_2$H$^+$)) is
found to be a good tracer for the evolutionary states of PGCC cores.
 Gas-grain chemical model can reproduce the
decreasing trend of $N$(C$_2$H)/$N$(N$_2$H$^+$) as a function of time.
The cores with the lowest  abundances of N$_2$H$^+$ ($X$[N$_2$H$^+$] $<$ $10^{-10}$) are the youngest, and have nearly constant abundances of C$_2$H.
In evolved cores with  $X$[N$_2$H$^+$] $\sim 10^{-9}$, abundances of C$_2$H drop quickly as the exhaustion of carbon atoms.
Although these PGCC cores are in different evolutionary states, they are all quite young  ($<$$5\times 10^5$ yr) with $N$(C$_2$H) $>$ $N$(N$_2$H$^+$).
Mapping observations are carried out toward 20 PGCC cores.
The PGCC cores in Cepheus have lower $N$(C$_2$H)/$N$(N$_2$H$^+$) and larger line widths compared with those in Taurus.
This  implies that  PGCC cores in Taurus are less chemically evolved than those in Cepheus.
\end{abstract}

\keywords{ISM: molecules -- ISM: abundances -- ISM: kinematics and dynamics  -- ISM: clouds -- stars: formation}

\clearpage

\section{Introduction}\label{intro}
The Planck satellite \citep{2010A&A...520A...2T, 2011A&A...536A...1P}
carried out the first all sky survey
in the submillimeter to millimeter range with unprecedented
sensitivity and provides a catalog of cold clumps of
interstellar matter in the Galaxy. The  Cold Clump
Catalog of Planck Objects (C3POs) released by \citet{2011A&A...536A..23P} consists of 10\,342 cold sources that stand out
against a warmer environment. The C3PO clumps are cold with
dust temperatures ranging from 7 to 19 K, peaking around 13 K.
Among the C3PO clumps, 915 early cold cores (ECCs)
were identified with most valid detection and lowest dust temperatures ($<$15K).
\citet{2016A&A...594A..28P} released 13,188 Planck
Catalog of Galactic cold clumps (PGCCs) as the full version
of the ECC catalog.
The characteristics of coldness and quiescence make them good targets to investigate
the initial conditions of star formation, including both dynamic processes
and chemical states
\citep{2010A&A...518L..93J,2012A&A...541A..12J,2015A&A...584A..93J,2015A&A...577A..69P,2017ApJS..228...12T,2018A&A...614A..83J}.

\begin{figure*}[!htb]
\centering
\includegraphics[width=0.99\linewidth]{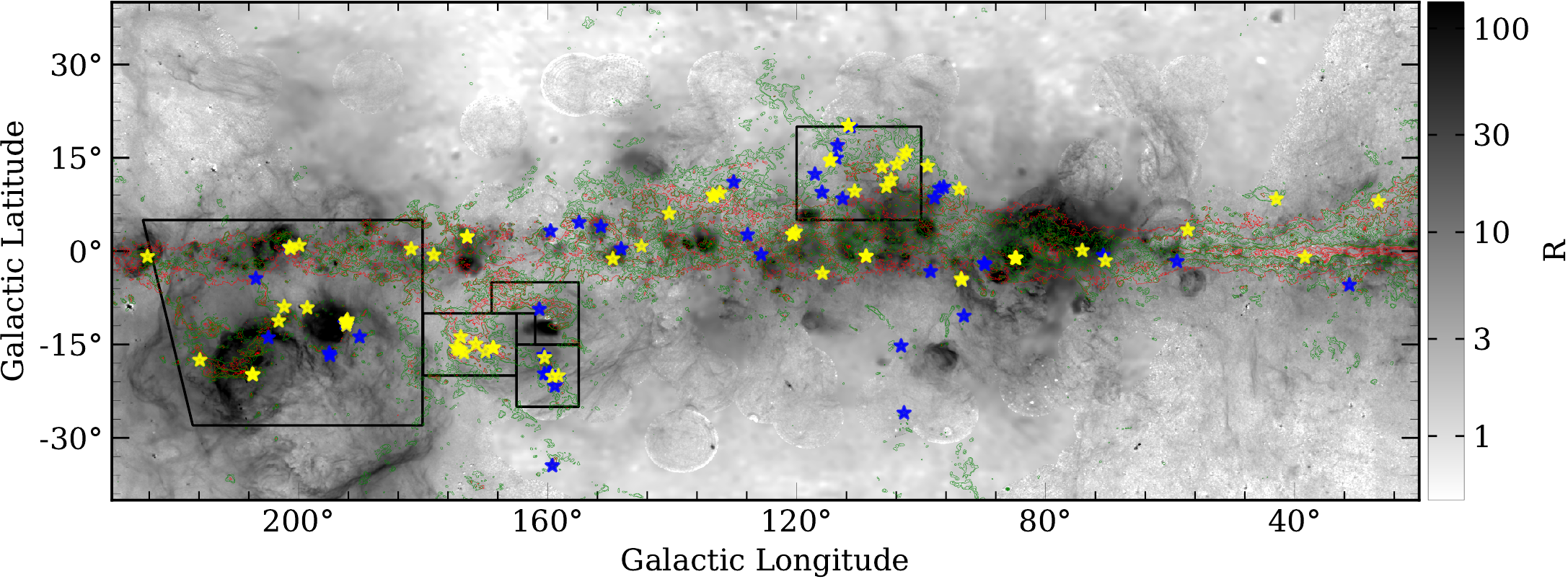}
\caption{Spatial distribution in the Galactic plane  of observed sources.
The CO-selected cores with and without detections of \CCH~are denoted by the
yellow and blue stars, respectively.
The background image represents the H$_{\alpha}$ emission \citep{2003ApJS..146..407F}
in unit of R (10$^6$/4$\pi$ photons cm$^{-2}$ s$^{-1}$ sr$^{-1}$).
The green contours represent CO (1-0) integrated emission detected by Planck HFI  \citep{2014A&A...571A..13P}.
The red contours show the Planck 353 $\mu$m continuum emission. The contour levels are (0.05, 0.1, 0.3, 0.5, 0.7, 0.9) $\times$
maximum value.
Famous star forming regions such as Taurus-Perseus-California \citep{2010A&A...512A..67L}, Cepheus, Orion complex
\citep{2001ApJ...547..792D} are  sketched with the black line.
\label{galac_dist}
}
\end{figure*}

Soon after the release of ECC data,  surveys with different molecular spectral lines were
conducted. Observations with the $J=1-0$ transitions of $^{12}$CO, $^{13}$CO,
and C$^{18}$O toward 674 Planck cold clumps selected
from the ECC catalog were performed by \citet{2012ApJ...756...76W}
using the 13.7 m telescope of the Purple Mountain Observatory (PMO).
Mapping observations of the same transitions were followed up soon
\citep{2012ApJS..202....4L,2013ApJ...775L...2L,2013ApJS..209...37M,
2015PKAS...30...79L,2016ApJS..222....7L,2016ApJS..224...43Z,2018ApJS..234...28L,
2018ApJ...859..151L,2018ApJS..236...49Z,2018ApJ...856..141T}.
Meanwhile, single-point observations of HCO$^+$ $J=1-0$ and HCN $J=1-0$  toward 621
CO-selected cores associated with PGCCs were performed \citep{2016ApJ...820...37Y}.
Thanks to these follow-up studies, the morphologies and dynamic properties
of PGCCs are fairly well understood. However, their locations in the evolutionary sequence are still unclear.
The chemical properties,  essential for understanding the evolutionary states of PGCCs, were not given
enough attention.
To investigate the chemical
evolutionary states of PGCCs, probing a large sample of such sources with
molecule pair of the early formed molecule ethynyl radical
(\CCH) \citep{2008ApJ...675L..33B} and the daughter molecule diazenylium (\NNHp) \citep{2003ApJ...593..906A,2017ApJS..228...12T}
will be helpful.

\CCH~is the simplest hydrocarbon molecule
with the carbon-carbon triple bond (C$\equiv$C).
Since being firstly detected by  \citet{1974ApJ...193L.115T},
\CCH~is found to be widely distributed and detected in all evolutionary stages of star-forming
regions \citep{2013ApJ...773..123S,2015ApJ...808..114J}.
\citet{2008ApJ...675L..33B} suggest that this molecule could also be used to study the
cold gas of forming stars to investigate their initial conditions.
Meanwhile, \NNHp~is also  an excellent tracer of dense molecular cloud cores \citep{2002ApJ...572..238C}.
\NNHp~is durable in cold  and dense regions owing to the depletions of its destroyers such as CO and the delayed freeze-out of its precursors such as  N$_2$.
We expect that \CCH~and \NNHp~are enhanced in different evolutionary states of PGCCs.

In this paper, we report a survey of \CCH~$N=1-0$ and \NNHp~$J=1-0$ toward the gas structures enclosed by emission of $^{13}$CO $J=1-0$ in the PGCCs
\citep{2012ApJS..202....4L,2013ApJS..209...37M,2016ApJS..224...43Z}.
We have compared the spectra of our detected species
with those of CO and its isotopologues as well as HCO$^+$ and HCN \citep{2012ApJ...756...76W,2016ApJ...820...37Y} to reveal the characteristics of \CCH~and \NNHp in PGCCs.
We also have compared the abundances of \CCH~and \NNHp~with those predicted
by gas-grain chemical model to investigate the evolutionary states of single PGCC and PGCCs in different regions.
This paper is arranged as follows. We present a description of the
sample and observations in Sect. 2. The results of the molecular line observations are
presented in Sect. 3. We discuss the properties of these two species
and the chemical evolutionary states of detected sources in Sect. 4.
We summarize the paper in Sect. 5.

\begin{table*}
\small
\centering
\caption{Line parameters.\label{tab:freq}}
\begin{tabular}{ccccc}
\hline\hline
Species & Transition                     & Freq (GHz)    & $S_{ij}\mu^2$ (D$^2$) & $E_{up}$ (K)\\
\hline
\CCH~	&	$N=1-0$, $J=3/2-1/2$, $F= 1-1$	&	87.28415	&	0.14	&	4.191	\\
	&	$N=1-0$, $J=3/2-1/2$, $F= 2- 1$     &	87.31692	&	1.42	&	4.192	\\
	&	$N= 1-0$, $J=3/2-1/2$, $F= 1- 0$	&	87.32862	&	0.71	&	4.191	\\
	&	$N= 1- 0$, $J=1/2-1/2$, $F= 1- 1$	&	87.402	&	0.71	&	4.196	\\
	&	$N= 1- 0$, $J=1/2-1/2$, $F= 0- 1$	&	87.40716	&	0.28	&	4.197	\\
	&	$N= 1- 0$, $J=1/2-1/2$, $F= 1- 0$	&	87.44651	&	0.15	&	4.196	\\
\NNHp~&	$J= 1 - 0$, $F1= 1- 1$	&	93.17188	&	37.2	&	4.471	\\
	&	$J= 1 - 0$, $F1= 2- 1$	&	93.1737	&	62.0	&	4.471	\\
	&	$J= 1 - 0$, $F1= 0- 1$	&	93.17613	&	12.4	&	4.471\\
\hline
\end{tabular}
\end{table*}

\section{Sample and observation}
\subsection{Sample characteristics}

A sample consisting of 121 CO-selected cores with strongest emission of  $^{13}$CO $J=1-0$ \citep{2012ApJ...756...76W}
was selected to be observed in \CCH~$N=1-0$ and \NNHp~$J=1-0$.
Spectra of $J=1-0$ of CO, $^{13}$CO and C$^{18}$O at the center of observed cores were extracted from previous mapping observations
\citep{2012ApJS..202....4L,2013ApJS..209...37M,2016ApJS..224...43Z}.
The preliminary work of deriving line parameters from these CO data was done as described in \citet{2016ApJ...820...37Y}.
Basic information about our detected sources including their equatorial coordinates, distances,
H$_2$ column densities
of host PGCCs derived from dust continuum ($N^d$(H$_2$)) \citep{2016A&A...594A..28P}, and CO parameters
are listed in \autoref{tab:sources}.

Distances of these
sources are adopted from the literature \citep{2012ApJ...756...76W,2016A&A...594A..28P}.
For sources with no available distances in the literature, distances are adopted as
the values with the highest probabilities given by a Bayesian distance calculator
\citep{2016ApJ...823...77R}.
The distances
calculated by the Bayesian distance calculator are on average  30 percent
higher than those adopted from the literature (\autoref{tab:sources}).
\autoref{galac_dist} shows the spatial distribution of the observed sources.
These sources are biased toward nearby star-forming regions while
the Galactic plane is under-represented.
These properties are inherited from the  whole sample of ECCs and CO-selected cores
\citep{2012ApJ...756...76W,2016ApJ...820...37Y}.
The red and green contours represent Planck 353 $\mu$m continuum emission and Planck CO $J=1-0$ emission detected
by the High Frequency Instrument on the Planck Satellite (Planck HFI) \citep{2014A&A...571A..13P}, respectively.
Distribution of Planck 353 $\mu$m continuum is well correlated with  that of CO $J=1-0$ detected by Planck HFI.
The Planck HFI CO emission traces relatively dense regions on Galactic scale,
and our CO-selected cores tend to locate at the margins of these regions.

The excitation temperatures of CO $J=1-0$ (T$_{ex}$(CO)) for our CO-selected cores
range from 9 K to 21 K. The mean value of T$_{ex}$(CO) is 14 K with a standard error of 0.3 K,
and it is slightly larger than the value in \cite{2012ApJ...756...76W} and the  average dust temperature
(13 K) for C3POs  \citep{2011A&A...536A..23P}.
Sources in our sample generally have higher H$_2$ column densities than those of CO cores in other PGCC samples.
The H$_2$ column densities of CO cores in nearby star-forming regions range from 1 $\times$ 10$^{21}$ cm$^{-2}$ to
10 $\times$ 10$^{21}$ cm$^{-2}$ with a mean value of 2.2 $\times$ 10$^{21}$ cm$^{-2}$\citep{2013ApJS..209...37M},
and those in the Galactic second quadrant range from 0.6 $\times$ 10$^{21}$ cm$^{-2}$ to 36 $\times$ 10$^{21}$ cm$^{-2}$
with a mean value of 8 $\times$ 10$^{21}$ cm$^{-2}$ \citep{2016ApJS..224...43Z}.
The H$_2$ column densities ($N_{CO}$(H$_2$)) of our CO-selected cores are derived from $N$($^{13}$CO)
adopting the $^{12}$C/$^{13}$C isotope ratio    and
CO abundance ($X$[CO])   as the values in the solar neighbor with a Galactocentric distance $\sim$8 kpc
\citep{2013A&A...554A.103P,1994ARA&A..32..191W}, 63 and 9$\times$10$^{-5}$, respectively.
$N_{CO}$(H$_2$) cover the range of (3-70) $\times$ 10$^{21}$ cm$^{-2}$ with a mean value of 2.2 $\times$ 10$^{22}$ cm$^{-2}$.

Futhermore, 20 cores with valid detection of \CCH~$N=1-0$ and \NNHp~$J=1-0$ were selected to perform mapping
observations.

\begin{figure*}[!htb]
\centering
\includegraphics[width=0.46\linewidth,height=0.32\linewidth]{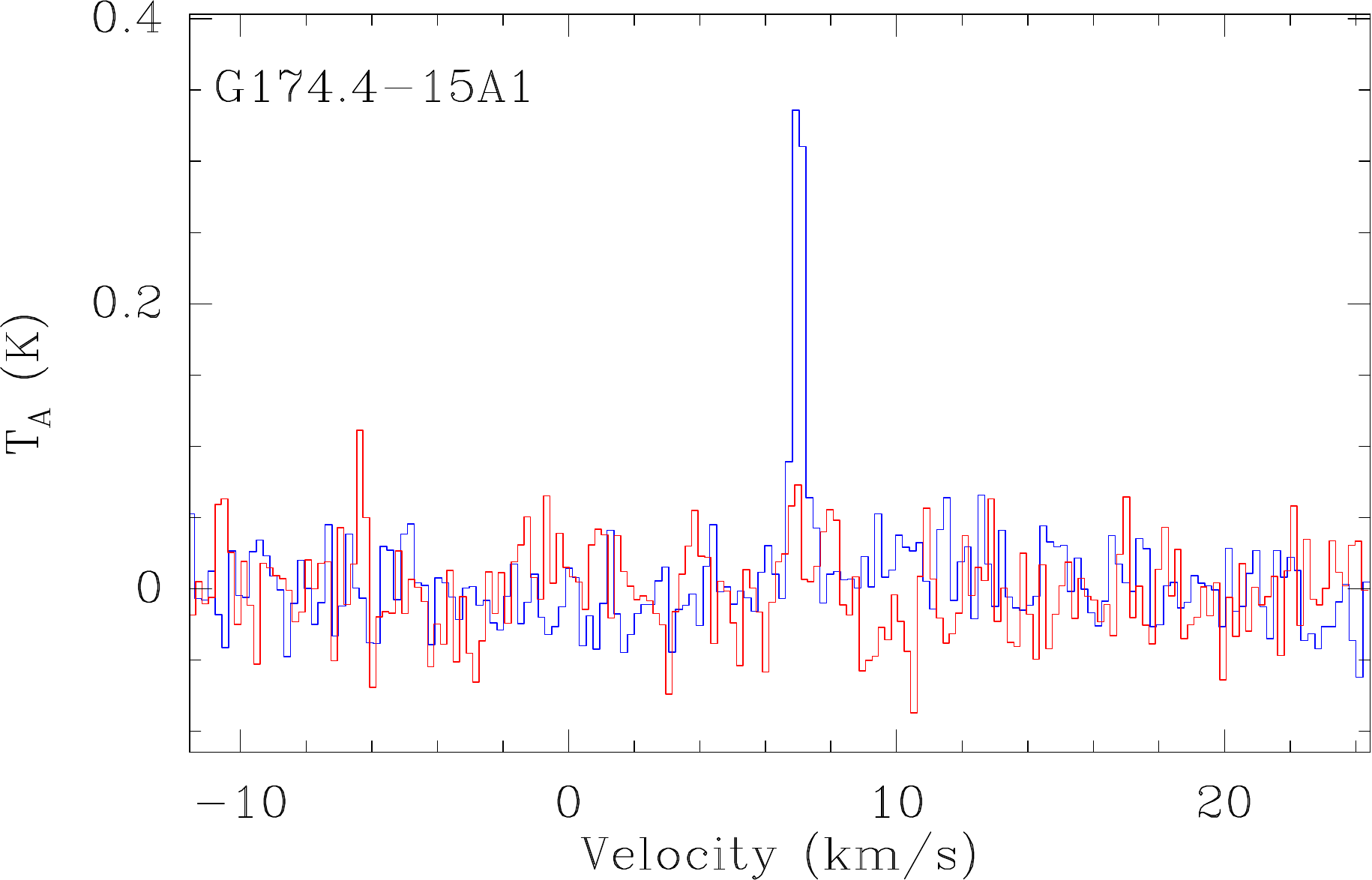}
\includegraphics[width=0.46\linewidth,height=0.32\linewidth]{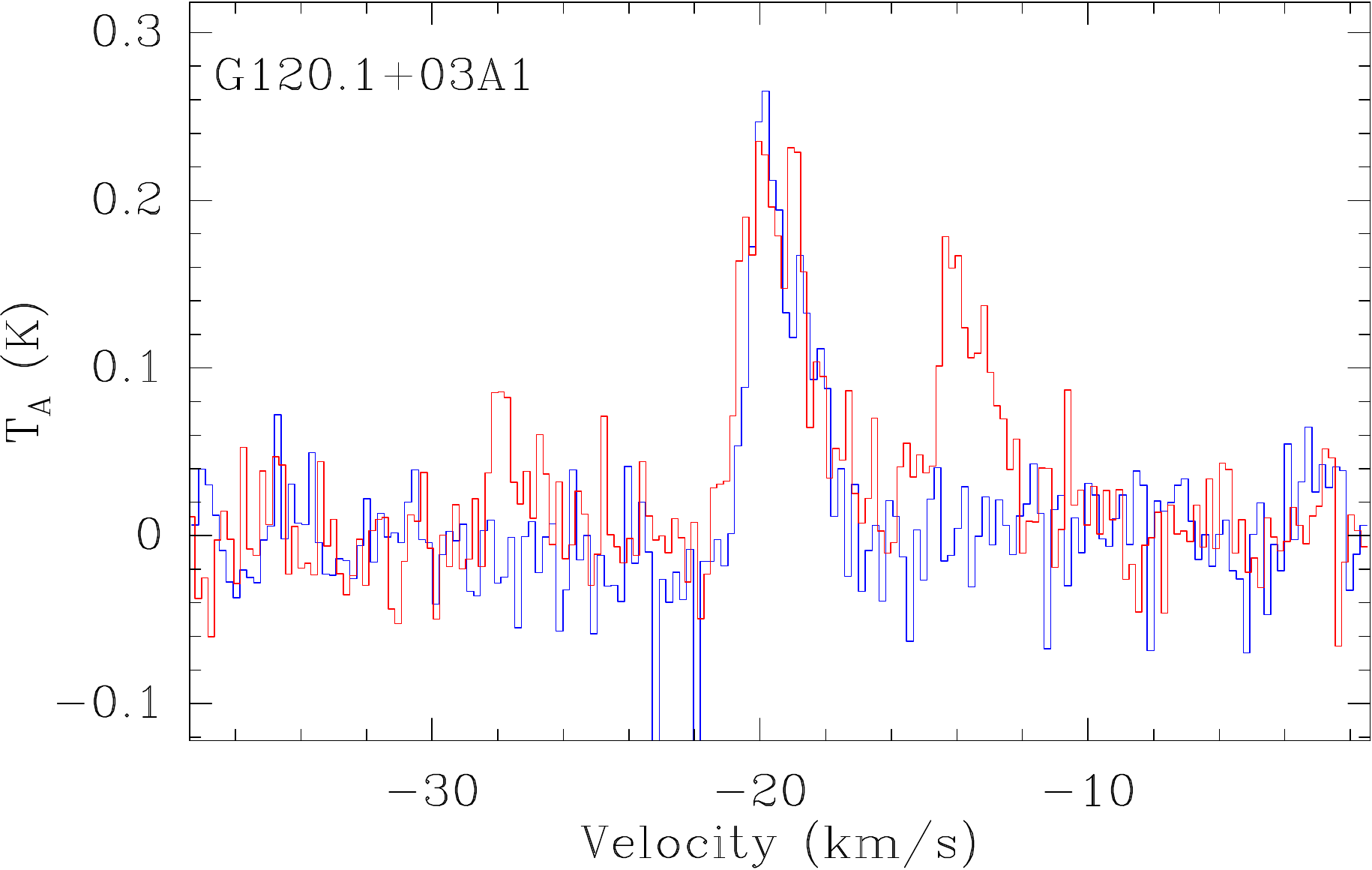}
\includegraphics[width=0.46\linewidth,height=0.32\linewidth]{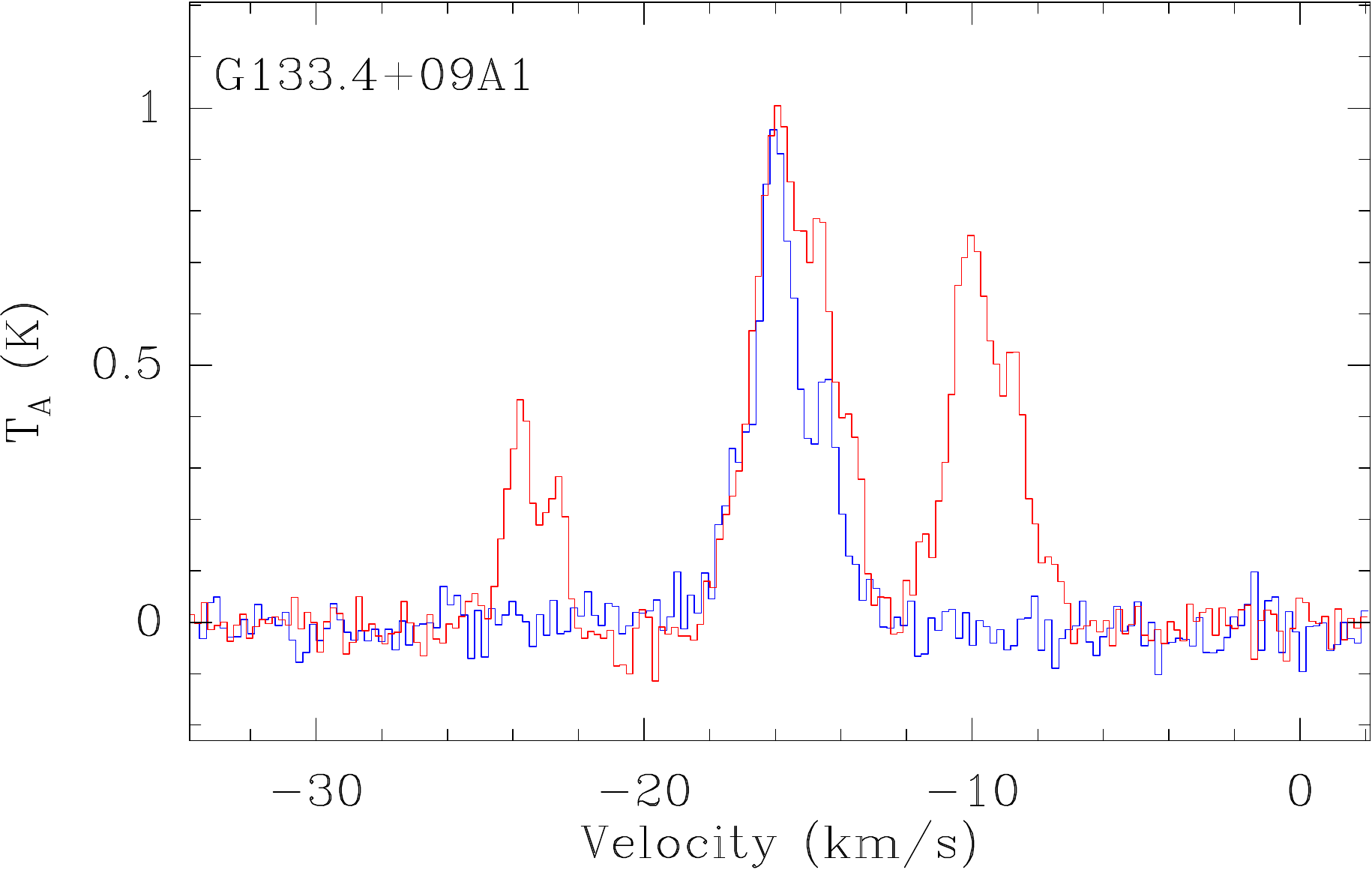}
\includegraphics[width=0.46\linewidth,height=0.32\linewidth]{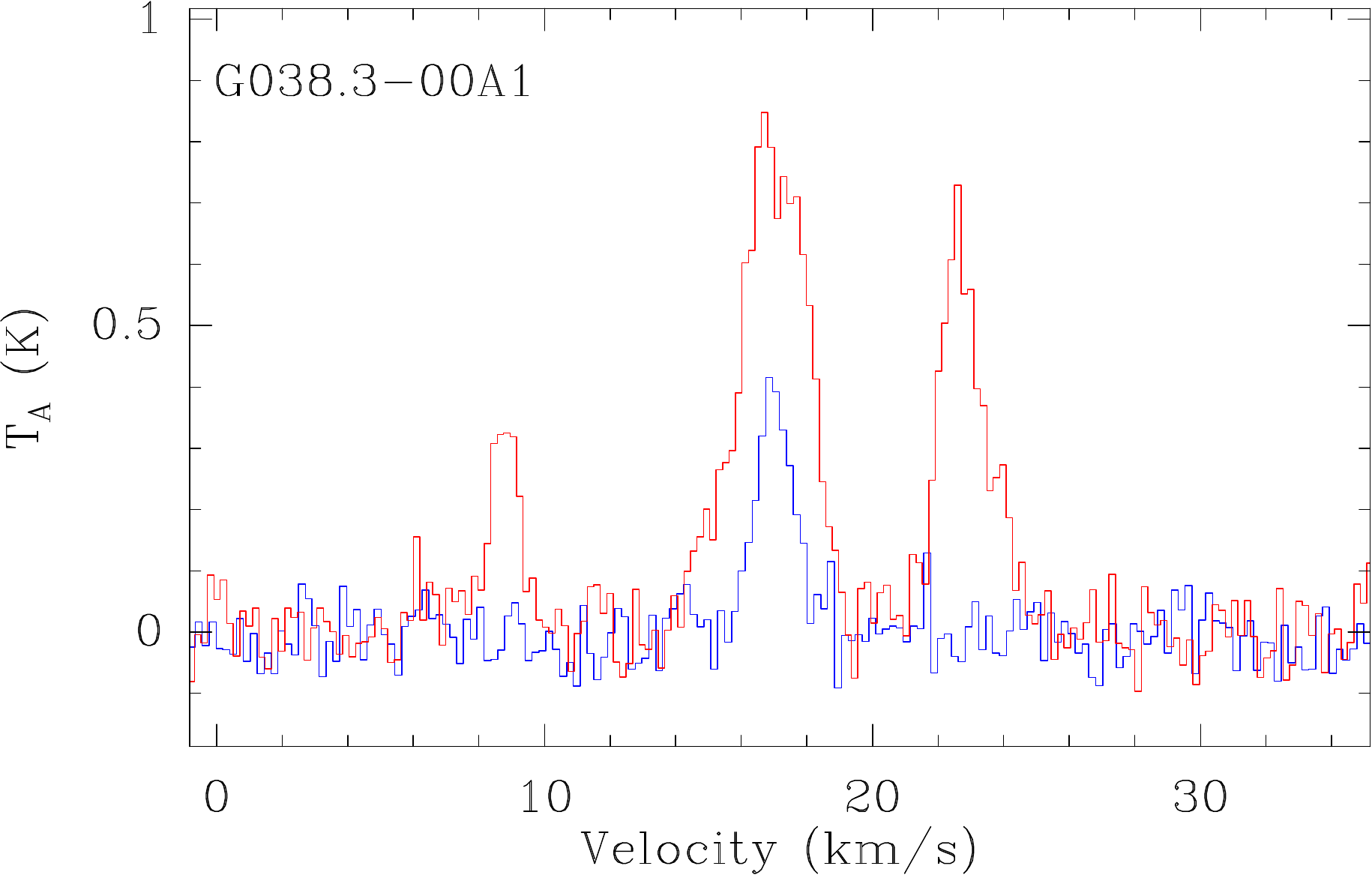}
\caption{Example spectra  of\ CCH~(blue) and \NNHp~(red)\label{example_spectra}.}
\end{figure*}

\subsection{Observations}
Single-point observations of \CCH~$N=1-0$ and \NNHp~$J=1-0$
(\autoref{tab:freq}) were carried out with
the PMO 13.7 m telescope from 2015 May to June.
The nine-beam Superconducting
Spectroscopic Array Receiver (SSAR) was working as the front
end in sideband separation mode (see \citealt{2012ITTST...2..593S}). An FFTS spectrometer was
used as back end, which has a total bandwidth of 1 GHz
and 16\,384 channels, corresponding to a velocity resolution
of 0.21 km s$^{-1}$ for \CCH~$N=1-0$ and 0.20 km s$^{-1}$ for \NNHp~
$J=1-0$. \CCH~$N=1-0$ was observed in the lower
sideband (LSB), while \NNHp~$J=1-0$ was observed
simultaneously in the upper sideband (USB). The half-power beam
width and main beam efficiency at 90 GHz are about 56\arcsec~and 0.5, respectively. The pointing
accuracy of the telescope was better than 4\arcsec. The typical system
temperature (T$_{sys}$) is around 170 K and varies about
ten percent. Spectra of \CCH~$N=1-0$ and  \NNHp~$J=1-0$ were integrated till
the rms of T$_{a}$  ranged from 20 mK to 50 mK.

Mapping observations  were performed in June 2015 using the PMO 13.7 m telescope.
Same front and back ends were employed as in single-point observations.
The on-the-fly (OTF) observation mode was applied. The antenna continuously scanned a region
of 18\arcmin~$\times$ 18\arcmin~centered on CO-selected cores with a scan speed of 20\arcsec~s$^{-1}$.
Only the central 10\arcmin~$\times$ 10\arcmin~regions were cut out for
further analyses because the edges of the OTF maps are very noisy. Data were
meshed with a grid spacing of 30\arcsec.

The GILDAS\footnote{\url{http://www.iram.fr/IRAMFR/GILDAS/doc/html/class-html}} package
including CLASS and GREG \citep{2000ASPC..217..299G, 2005sf2a.conf..721P}
was used to reduce the data. All figures were plotted using the open source Python package, Matplotlib.

\section{Results}
Among the 121 observed CO-selected molecular cores, 71 have detection of \CCH~$N=1-0$ and 58
have detection of \NNHp~$J=1-0$.
Cores with or without detection of \CCH~$N=1-0$ and \NNHp~$J=1-0$
are denoted in Figure 1 with stars in different
colors, and their projected spatial distributions have no obvious deviations.
Typical spectra of several cores with antenna temperature (T$_a$)
of \CCH~$N=1-0$, $J=3/2-1/2$, $F=2-1$ larger or smaller than or comparable with
that of \NNHp~$F1=2-1$ are shown in \autoref{example_spectra}  as examples.

Mapping observations of \CCH~$N=1-0$ and \NNHp~$J=1-0$ are performed toward 20 cores. Both lines are detected in all sources
except \NNHp~$J=1-0$ in G167.2-15A1.

\subsection{Line parameters}\label{chapter:line}
All six hyperfine structure (HFS) components  are well
resolved for  \CCH~$N=1-0$ of detected sources, while only  the spectra of the main component \CCH~$N=1-0$, $J=3/2-1/2$, $F= 2-1$ are exhibited in \autoref{example_spectra}.
However, only three groups of HFS components of \NNHp~$J=1-0$, as listed in \autoref{tab:freq},
are well resolved for  most of the sources.
$F1=2-1$ is the main group
of \NNHp~$J=1-0$ and consists of two hyperfine lines labeled as $F=2-1$ and $F=3-2$ with a
velocity separation  $\sim$1 km s$^{-1}$.

Using the HFS fitting program in
GILDAS/CLASS, we performed hyperfine structure fitting
toward spectra of \CCH~$N=1-0$ and \NNHp~$J=1-0$.
In the HFS fitting, the optical depths of different hyperfine lines are all assumed  as Gaussian with
the same width, and
the excitation temperatures for different HFS components are the same \citep{2016A&A...592A..21F}.
Hyperfine structure fitting can give the parameters such as line width ($\Delta$V)  and velocity (V$_{\mathrm{LSR}}$) very precisely.
The results of HFS fittings  are listed in \autoref{tab:basic},
including  T$_a$, V$_{\mathrm{LSR}}$, $\Delta$V,
and integrated intensities ($\int T_ad V$) of \CCH~$N=1-0$, $J=3/2-1/2$, $F=2-1$ and \NNHp~$J=1-0$, $F1=2-1$.
The optical depths are not listed because most of the lines we detect are optical thin ($\tau$ $<$ 0.1),
and hence HFS fitting can not
provide accurate values of optical depths.

The T$_a$ of \CCH~$N=1-0$, $J=3/2-1/2$, $F=2-1$ ranges from 0.08 K to 0.93 K with a median value of 0.50 K.
The T$_a$ of \NNHp~$N=1-0$,  $J=1-0$, $F1=2-1$ ranges from 0.10 K to 1.03 K with a median value of 0.45 K.
 The sources with detection of \NNHp~$J=1-0$ all have detection of \CCH~$N=1-0$.
Only 10 sources
have  T$_a$ of \NNHp~$J=1-0$, $F1=2-1$  larger than that of \CCH~$N=1-0$, $J=3/2-1/2$, $F=2-1$ by
more than 3 $\sigma$  (\autoref{tab:basic}), and all of them have
line widths of \CCH~$N=1-0$ and C$^{18}$O larger than the average
line width of \CCH~($\sim$1.0 km s$^{-1}$) except G104.4+06A1.

\begin{figure*}[!hbt]
\centering
\includegraphics[width=0.46\linewidth,height=0.35\linewidth]{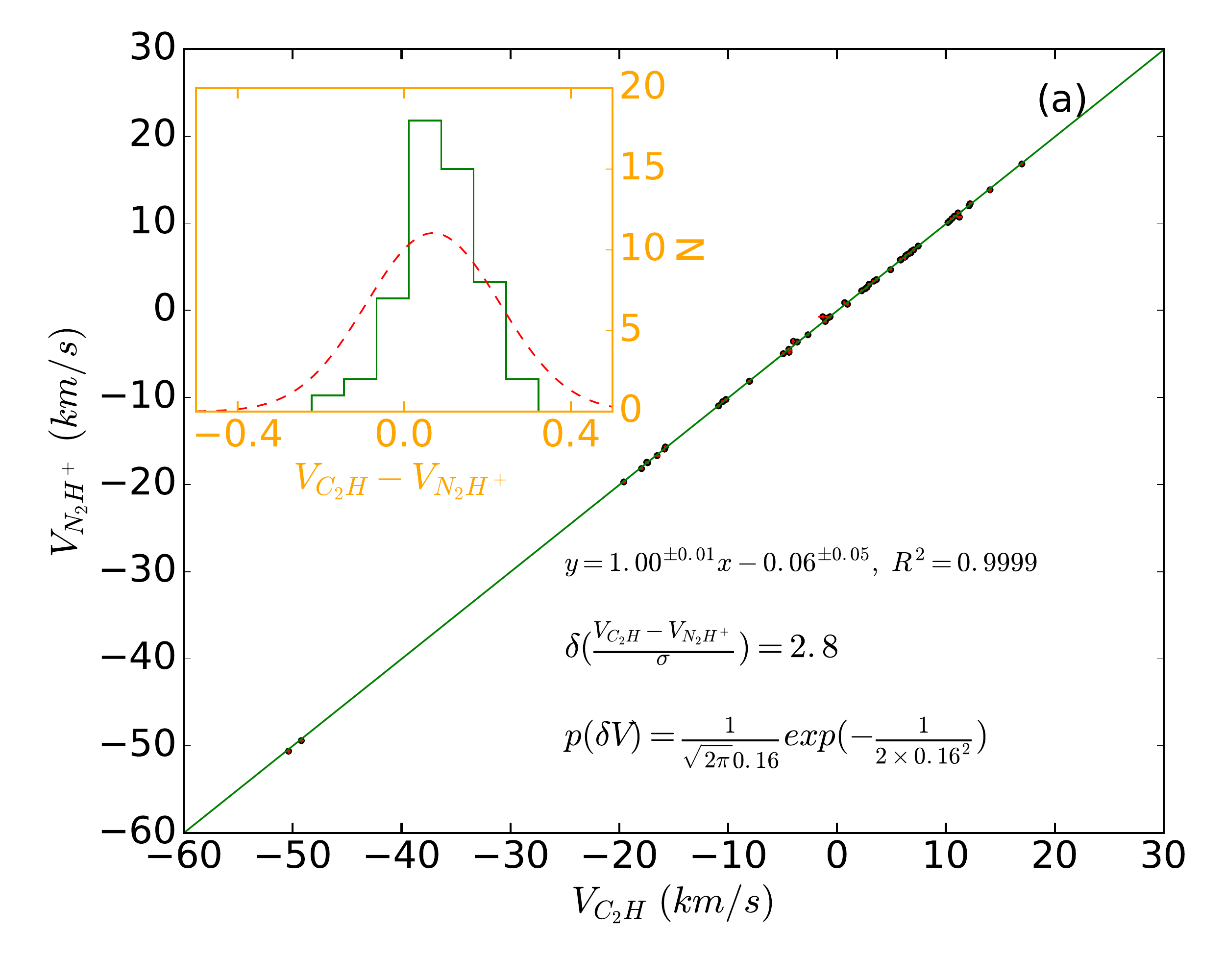}
\includegraphics[width=0.46\linewidth,height=0.35\linewidth]{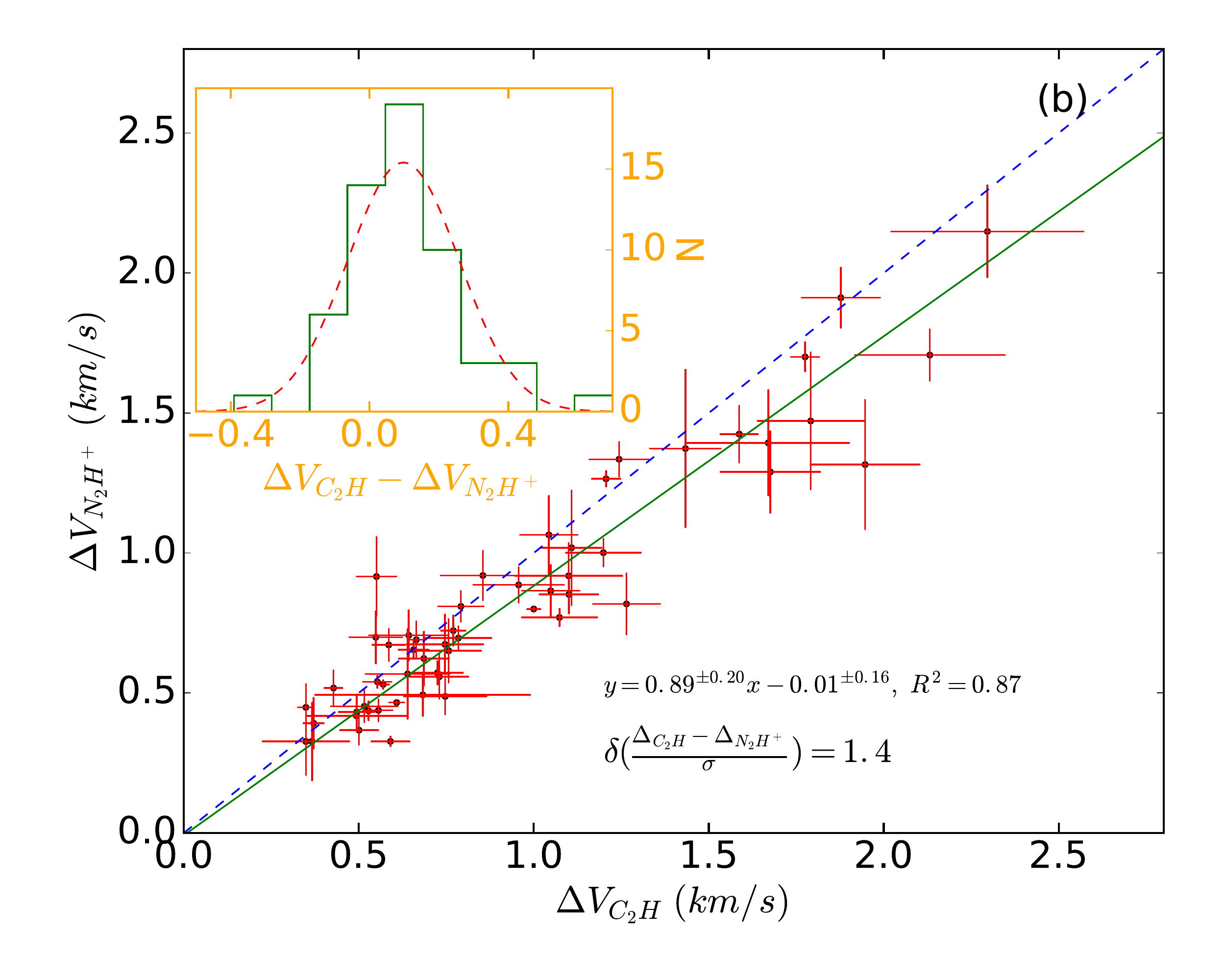}
\includegraphics[width=0.46\linewidth,height=0.35\linewidth]{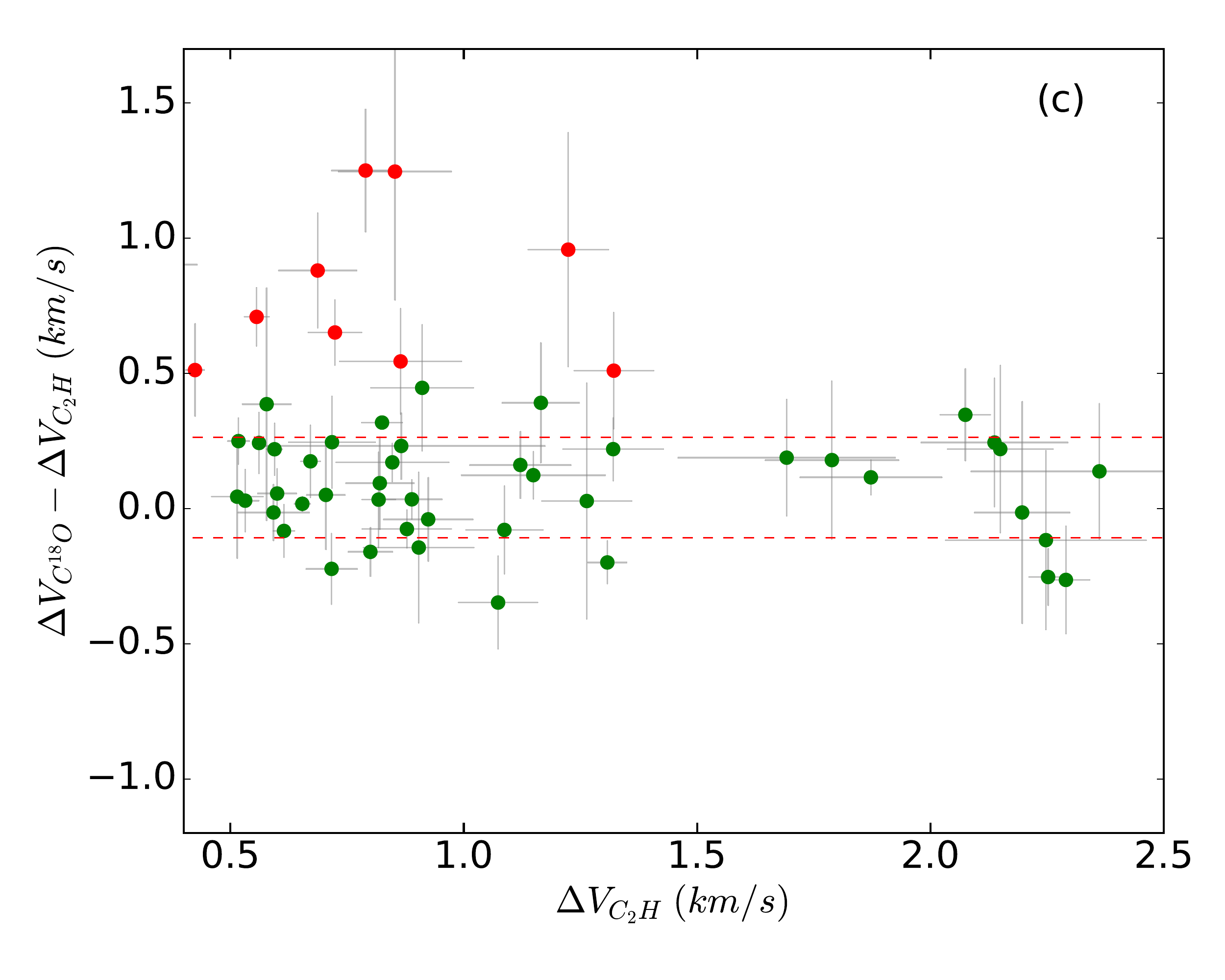}
\includegraphics[width=0.46\linewidth,height=0.35\linewidth]{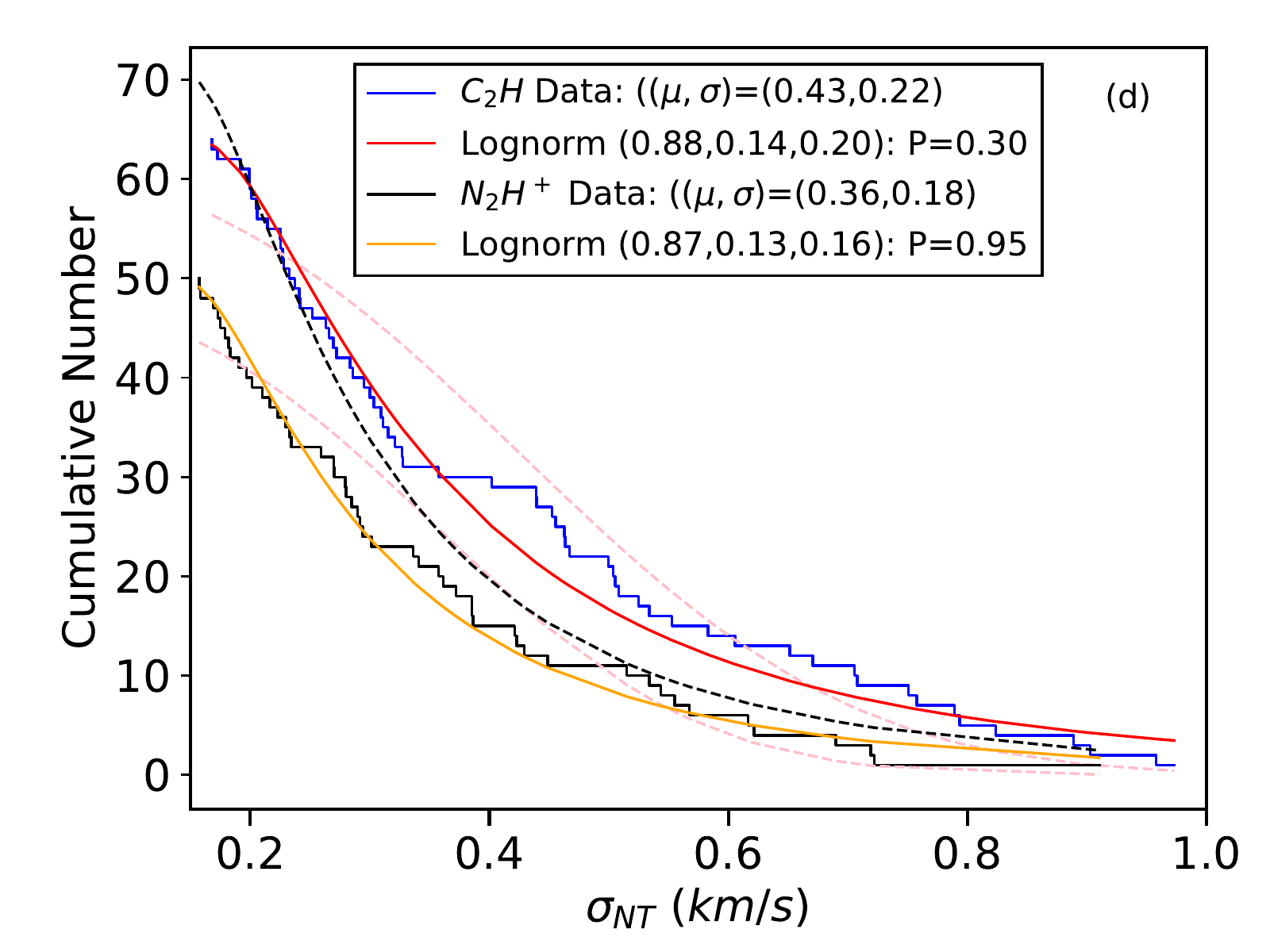}
\caption{\label{kinematics}
Panel (a): Correlation between the centroid velocity  of \CCH~$N=1-0$ and that of \NNHp~$J=1-0$.
The green solid line represents the result of linear least-squares fitting.
Panel (b): Correlation between the line width of \CCH~$N=1-0$ and that of \NNHp~$J=1-0$.
Blue dashed line represents $\Delta$V$_{C_2H}$= $\Delta$V$_{N_2H^+}$.
Panel (c): Correlation between line width of \CCH~$N=1-0$ and the difference between the line width of \CCH~$N=1-0$ and that of C$^{18}$O.
The two dashed lines denote the 1 $\sigma$ values for the distribution of line width differences.
Panel (d): Fittings of the cumulative distribution functions of nonthermal velocities.
Red and yellow solid lines show results of lognormal fittings. 
The three parameters (\autoref{eq_log}) of the best lognormal fits are
(0.88, 0.14, 0.20) and (0.87, 0.13, 0.16), respectively.
Dashed pink lines show results of standard-normal fittings.
}
\end{figure*}

\autoref{kinematics}(a) shows the correlation between the V$_{\mathrm{LSR}}$ of \CCH~$N=1-0$ and \NNHp~$J=1-0$.
They agree with each other very well, with  $\delta((V_{C_2H}-V_{N_2H^+})/\sigma_1)$ smaller than three,
where $\sigma_1^2=\sigma(V_{C_2H})^2+\sigma(V_{N_2H^+})^2$.
From \autoref{kinematics}(b), one can see that
line widths of \CCH~$N=1-0$ and \NNHp~$J=1-0$ are also quite consistent with each other and
$\delta$(($\Delta_{C_2H}$-$\Delta_{N_2H^+})/\sigma_2$) is smaller than 1.5,
where $\sigma_2^2=\sigma(\Delta_{C_2H})^2+\sigma(\Delta_{N_2H^+})^2$.
The mean widths of $^{13}$CO $J=1-0$ and C$^{18}$O $J=1-0$ of these CO-selected cores with
valid detection of \CCH~$N=1-0$ are 2.0 and 1.3 km s$^{-1}$,
slightly larger than the mean width  of \CCH~$N=1-0$.
However, from \autoref{kinematics}(c) it can be clearly seen that the larger mean width of C$^{18}$O
mainly results from several sources (shown as red dots) with $\Delta$V$_{C_2H}$ smaller than
1 km s$^{-1}$.
The $\Delta$V of \CCH~$N=1-0$ are well consistent with those of C$^{18}$O $J=1-0$ for the rest of the sources, especially for sources
with widest line widths.
\autoref{kinematics}(d) shows the cumulative distribution functions of  nonthermal velocities (\autoref{eq_sigNT})
traced by \CCH~$N=1-0$ and \NNHp~$J=1-0$.
Nonthermal velocity dispersions, $\sigma_{NT}$, traced by \CCH~$N=1-0$ and \NNHp~$J=1-0$ can be better fitted with lognormal
distributions than Gaussian distributions.
The probability density function function of lognormally distributed random variable $X$ ($f_X$) can be expressed
with three parameters (a, b, c) as
\begin{equation}
f_X(x) dx =\frac{1}{\sqrt{2\pi}a} \exp\left(-\frac{\left(\ln(x-b)-\ln(c)\right)^2}{2a^2}\right)d\ln(x-b) \label{eq_log}
\end{equation}
The mean ($\mu$) and variance ($\sigma^2$) of lognormally distributed random variable $X$ can be expressed as
\begin{equation} \mu=E[X]=\exp(a^2/2)\times c+b \end{equation}
\begin{equation}\sigma^2=Var[X]={c^2\left(\exp(2a^2)-\exp(a^2)\right)}\end{equation}
The three parameters of the best lognormal fits are
(0.88, 0.14, 0.20) and (0.87, 0.13, 0.16) (\autoref{kinematics}(d)),
corresponding to mean values and standard deviations ($\mu$, $\sigma$) of the fitted curves
(0.43, 0.32) km s$^{-1}$  and (0.36, 0.25) km s$^{-1}$,
respectively.

\subsection{Derived parameters}\label{derived_par}
The dispersions of thermal velocity ($\sigma_{therm}$) and one dimensional nonthermal velocity ($\sigma_{NT}$)
 can be calculated as
\begin{equation} \sigma_{therm}  = \left[\frac{kT_{therm}}{m_H\mu_{H_2}}\right]^{\frac{1}{2}} \end{equation}
\begin{equation} \sigma_{NT}  = \left[\sigma_{}^2-\sigma_{therm}^2\frac{m_H\mu_{H_2}}{m_X}\right]^{{\frac{1}{2}}}\label{eq_sigNT}\end{equation}
where $\sigma=\Delta V/\sqrt{8\ln(2)}$, $T_{therm}$ is the gas kinetic temperature
which is adopted as excitation temperature of CO,
k the Boltzmann's constant, m$_X$ the
molecular mass, m$_H$  the mass of atomic hydrogen, and $\mu_{H_2}$ = $\rho$/n(H$_2$)  the mean molecular weight of the gas
\citep{2008A&A...487..993K}.
$\mu_{H_2}$ is adopted as 2.72 assuming n(He)/n(H)=0.18 and  ignoring the mass contributions of metals.
The $\sigma_{NT}$ derived from emission of \CCH~$N=1-0$ and that of \NNHp~$J=1-0$ are listed in the second and third columns of
\autoref{tab:derived}, respectively.

Under the assumption  of local thermal equilibrium (LTE), column densities of \CCH~and \NNHp~can be calculated
through \citep[e.g.][]{2015PASP..127..266M}
\begin{equation}
 N=\frac{3k}{8\pi^3\nu}\frac{Q}{\sum S_{ij}\mu^2}
exp\left(\frac{E_{up}}{kT_{ex}}\right) J(T_{ex})
  \int \tau d V \label{Eq_N}
\end{equation}
\begin{equation}T_{r,\nu}= \frac{h\nu}{k}[J(T_{ex}-J(T_{bg}))]\times [1-exp(-\tau_\nu)]f \label{Eq_tau} \end{equation}
where J(T)  = $[ exp ( h\nu/kT ) - 1 ]^{ -1}$, T$_\mathrm{bg}$ (2.73 K) is the
temperature of the cosmic background radiation,
h is the Planck constant, and the
beam-filling factor f is assumed as unit.
The permanent dipole moment  $\mu$, line strength S$_{ij}$, partition function Q and
upper level energy E$_u$
were adopted from the Cologne Database for Molecular Spectroscopy \footnote{\url{http://www.astro.uni-koeln.de/cdms/}}
and partly listed in \autoref{tab:freq}.

Unfortunately, the excitation temperatures cannot be given by HFS fittings
because most lines we detected are optical thin and the exact beam filling factors are unknown for single-point sources.
Besides, the assumption that the excitation temperatures of different  hyperfine components stay the same
is not always valid. For example, the differences among the excitation temperatures of different  hyperfine components of \CCH~$N=1-0$
can be as large as several K in L1498 \citep{2009A&A...505.1199P}.
Therefore, the excitation temperatures T$_{ex}\sim 5$ K were adopted.
For optical thin lines, the assumption, T$_{ex}$ = $E_u/k$, was usually made to give the lower limits to
the column densities \citep{2014A&A...562A...3M}.
It is also consistent with the typical excitation temperatures of the
\NNHp~$J=1-0$  (5 K) in dense cloud cores \citep{2002ApJ...572..238C}.
To explore how large uncertainties are brought in under this estimation,
we also calculated the column densities with excitation temperatures deduced from spectra of $J=1-0$ of
$^{12}$CO and $^{13}$CO.
The column densities of \CCH~and \NNHp~as well as their ratios calculated based  on the two set of T$_{ex}$ assumptions
are listed in the fourth-to-sixth and seventh-to-ninth columns of \autoref{tab:derived}, respectively.
It is clear from \autoref{tab:derived}  that column densities calculated based on the two set of T$_{ex}$ assumptions
do not deviate much from each other,
and most of them have deviations less than 15 percent.
An underestimation of T$_{ex}$ (5 K) would introduce a higher $\tau$ through \autoref{Eq_tau}, which compensates
for the reduced  column densities introduced by it in \autoref{Eq_N}.
The values of $N$(\CCH)/$N$(\NNHp) change little within a wide range of temperatures \citep{2017ApJ...836..194P}.
The column densities calculated with T$_{ex}$ = 5 K are adopted for further discussions.

\begin{figure*}[p]
\centering
\includegraphics[width=0.9\linewidth]{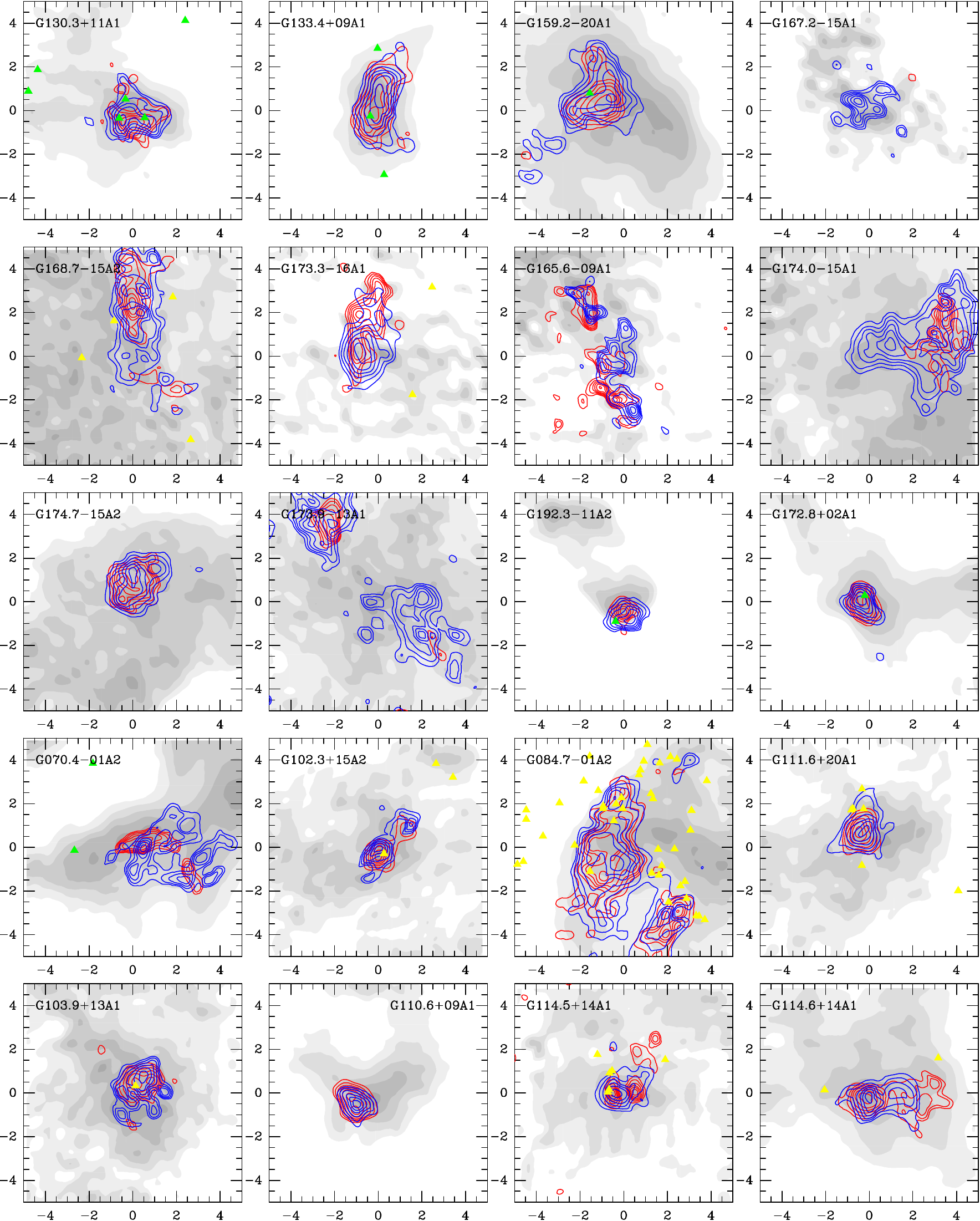}
\caption{Contours of integrated intensities of \CCH~$N=1-0$ (blue) and \NNHp~$J=1-0$ (red) from
45 percent to 95 percent stepped by 10 percent of maximum value.
Background shows $^{13}$CO emission. Yellow triangles and green stars represent 2MASS sources and IRAS sources
quoted from Simbad respectively. \label{fig_map}}
\end{figure*}

\subsection{Mapping parameters}
Among sources with detection of \CCH~$N=1-0$ and \NNHp~$J=1-0$,  20  were performed with  mapping observations,
and all have detection of \CCH~$N=1-0$ and \NNHp~$J=1-0$  with a $\sigma$ $\sim$ 0.08 K km s$^{-1}$ except G167.2-15A1.
The nondetection of \NNHp~$J=1-0$ for G167.2-15A1 might be due to its weak emission (Ta $\sim$ 0.33 K) and
shorter integration time ($\sigma\sim$ 0.13 K) compared with other mapped sources.
The integrated intensities of \CCH~$N=1-0$ and \NNHp~$J=1-0$ are shown as blue and red contours in \autoref{fig_map}, respectively.

From each map shown in \autoref{fig_map}, one or several substructures are resolved in one CO-select core.
The contour with half maximum value
is taken as the border line of a substructure.  Additional labels are used to distinguish different substructures
if there are two or more substructures resolved  within a single map, for example ``NE'' if one substructure was in the northeast
relative to its neighbor substructure.
Since the emission regions of \CCH~$N=1-0$ and \NNHp~$J=1-0$ are
well correlated with each other, each substructure resolved from \CCH~$N=1-0$ map is matched with
its nearest substructure resolved from \NNHp~$J=1-0$ map, and a same location label is given to them.
The radius ($r$) of a substructure is defined as the radius of a circle whose
area is equal to the area enclosed by the border line of that substructure.
In total there are 26 substructures resolved with $r$  $>$ 0.5\arcmin.
The average value of the radius for \CCH~substructures is 1.3$\pm$0.1 arcmin (0.28$\pm$0.05 pc),
and it is larger than that for \NNHp~substructures, 0.9$\pm$0.1 arcmin (0.22$\pm$0.05 pc).
The names of their harboring CO-selected cores and their location labels are listed in the first two columns  of \autoref{tab:map}.
Process of calculating the peak column densities
is the same as that in Sect. \ref{derived_par}.
Parameters including peak positions,
V$_{\mathrm{LSR}}$, $\Delta$V, and column densities of peak points as well as
the radii  of these substructures are also listed in \autoref{tab:map}.

Because the abundances of \CCH~are less variant than those of \NNHp,
the masses of substructures ($M_{sub}$)  are calculated based on emission  of \CCH~$N=1-0$ through the equation
\begin{equation} M_{sub} = \frac{\mu m_Hd^2}{X[C_2H]}\int N(C_2H)dS  \end{equation}
where the abundance of \CCH~($X$[\CCH]) is assumed as 10$^{-8}$.
It is reasonable because abundances of  \CCH~are nearly constant
in  diffuse molecular gas (4$\pm$2 $\times$ 10$^{-8}$) \citep{2008ApJ...675L..33B}, starless cores such as TMC-1 (3-5 $\times$ 10$^{-8}$) \citep{2018ApJ...856..151L} and
prestellar cores such as L1498 (0.8$\pm$0.1 $\times$ 10$^{-8}$)  \citep{2009A&A...505.1199P} as well as PGCCs
(Sect. \ref{abundance_evolution}).
The uncertainty of calculated $M_{sub}$ contributed by the assumption of fixed $X$[\CCH] can be as high as a factor of five.
Virial masses ($M_{vir}$) of dense cores assumed as gravitationally bounded spheres with $\rho\propto R^{-2}$ can be calculated though
\citep{1988ApJ...333..821M,1994ApJ...428..693W}
\begin{equation} M_{vir} = \frac{5R\sigma^2_{3D}}{3\gamma G}\end{equation}
where $\sigma_{3D}^2 = 3(\sigma_{NT}^2+\sigma_{therm}^2$), G is the gravitational constant, $\gamma=5/3$.
The $M_{sub}$, $M_{vir}$ and virial parameters $\alpha=$ $M_{vir}$/$M_{sub}$ are listed in the last three columns of \autoref{tab:map}.

The derived virial parameters range from 1.2 to 21.8 with a median value of 4.8.
Among the 26 substructures resolved, 20 have virial parameters smaller than five.
Considering the possible underestimations of  the masses of substructures for the overestimations of  $X$[\CCH],
most of these substructures in PGCCs are approximately virialized and slightly confined by external pressures \citep{2015MNRAS.450.1094P}.

\begin{figure*}[!hbt]
\centering
\includegraphics[width=0.46\linewidth,height=0.35\linewidth]{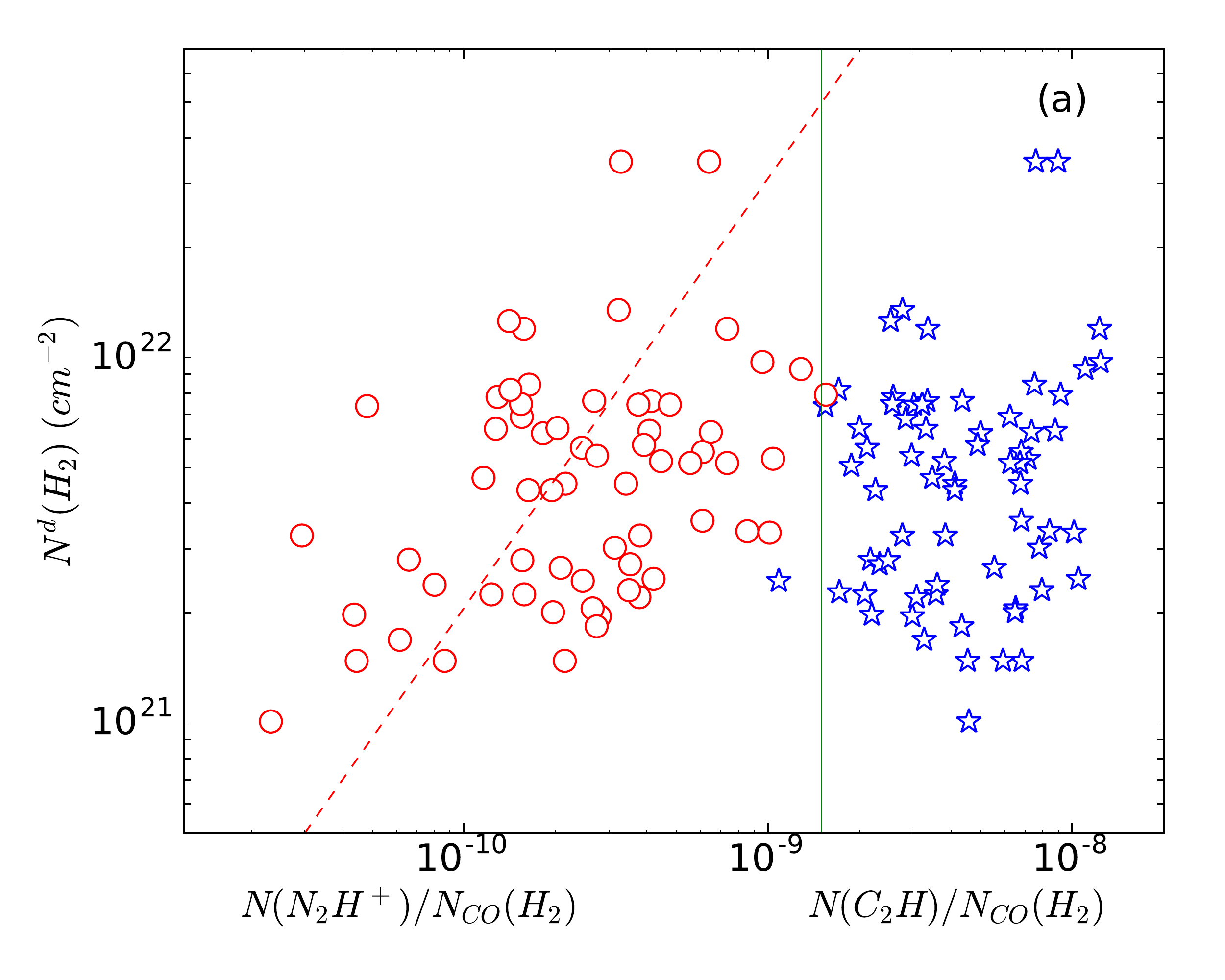}
\includegraphics[width=0.46\linewidth,height=0.35\linewidth]{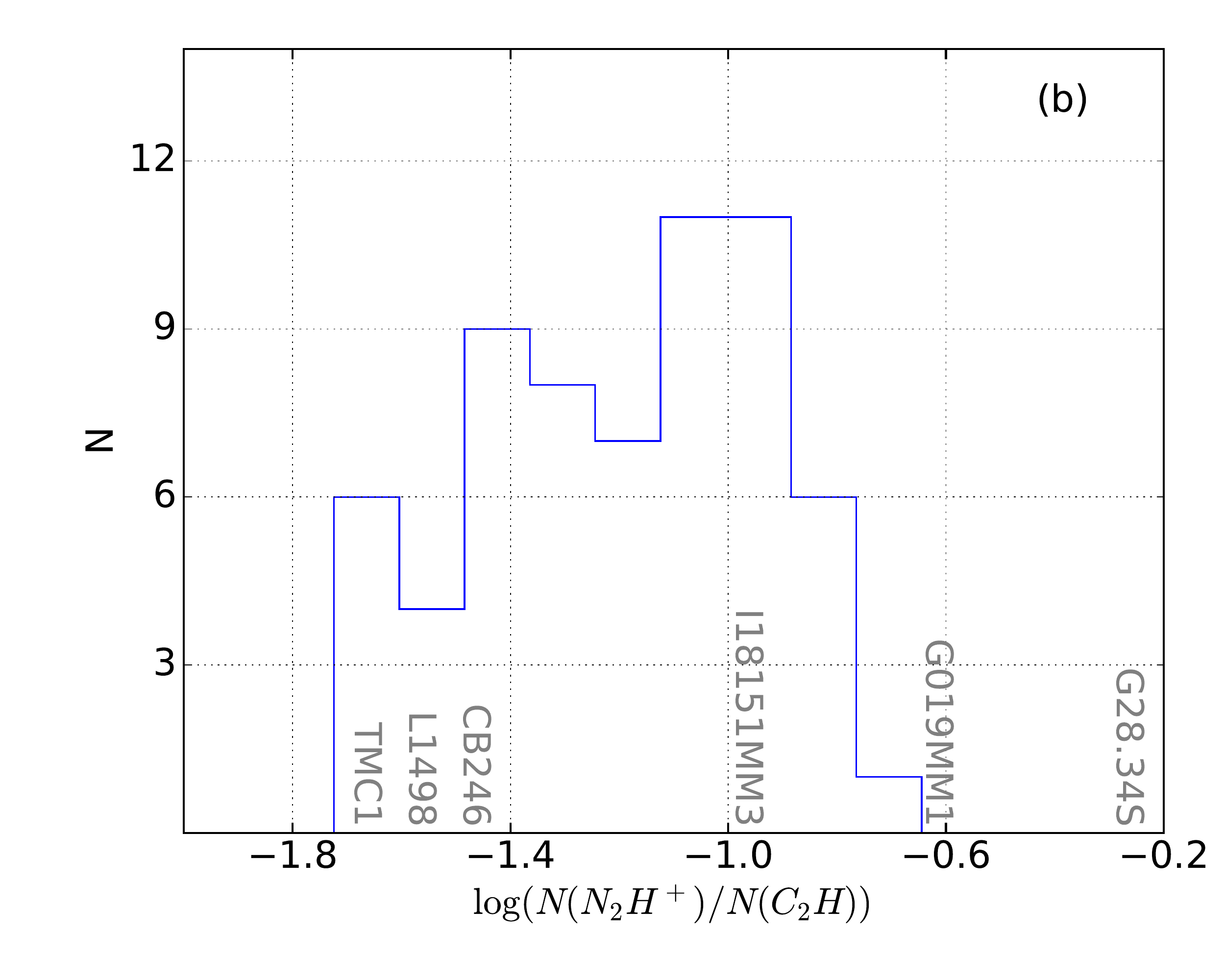}
\caption{\label{fig_relation}
Panel (a): Relation between the abundances  of \CCH~and \NNHp~of CO-selected cores
and the H$_2$ column densities of their host PGCCs ($N^d$(H$_2$)).
The abundances of \CCH~and
\NNHp~of our sources locate in two different regions separated by
the green line.
The dashed red line shows the result of linear least-square fitting on data represented by red circles.
Panel (b): Blue line shows number density distribution of log($N$(N$_2$H$^+$)/$N$(C$_2$H)) for our detected PGCC sources.
The value of this parameter for typical sources are also shown, including
well known starless cores TMC-1 \citep{2004ApJ...617..399H,2018ApJ...856..151L}, L1498 and
CB246 \citep{2009A&A...505.1199P},
massive clumps associated with infrared dark clouds I18151-1208 MM3
(abbreviated as I18151MM3) and G019.27+0.07 MM1 (abbreviated as G019MM1)
\citep{2010ApJ...714.1658S,2008ApJ...678.1049S}, as well as the infrared dark cloud G028.23-00.19 \citep{2013ApJ...773..123S}.
}
\end{figure*}

\section{Discussion}
\subsection{Kinematics}
All sources have only one velocity component with single peak, except for G120+03A1 and G133.4+09A1
whose spectra exhibit blue asymmetry.
A double-peaked line is blue or red asymmetric if its higher peak skews to the blue or red side.
Combining with an optically thin line,
an optically thick double-peaked line  can be further identified as a blue profile if its higher peak is shifted
blueward with $\delta$V = (V$_{thick}$-V$_{thin}$)/$\Delta$V$_{thin}$ $<$ -0.25 \citep{1996ApJ...465L.133M,1997ApJ...489..719M}.
A red profile would have $\delta$V $>$ 0.25 \citep{2007ApJ...669L..37W,2013MNRAS.429..954Y}.
However, the characteristic blue profile will appear only if the molecular tracer has a suitable optical depth and critical density
\citep{2003ApJ...592L..79W,2005ApJ...626..919E} within a source with warmer center regions \citep{1993ApJ...404..232Z}.
Blue profiles are not expected to be common in  PGCCs with dark and cold center regions.

Similar profiles may also be produced by multicomponents of target sources.
In G120+03A1,  spectrum of HCO$^+$ $J=1-0$ was identified as red profile (with higher red peak) by \citet{2016ApJ...820...37Y}.
Although emission of \NNHp~$J=1-0$ is weak and blended by several hyperfine structures, a
small blue component can still be resolved. Spectrum of C$^{18}$O $J=1-0$ has two peaks similar to
that of \CCH~$N=1-0$ while that of $^{13}$CO $J=1-0$ shows a flat-topped single peak  \citep{2016ApJ...820...37Y}.
In G133.4+09A1, both spectra of  \CCH~$N=1-0$ and \NNHp~$J=1-0$ show two resolvable peaks.
A small red component can also be resolved with C$^{18}$O $J=1-0$,  but not with  HCN $J=1-0$ and HCO$^+$ $J=1-0$ \citep{2016ApJ...820...37Y}.
Spectrum of HCO$^+$ $J=1-0$ shows broad line wings and there are multiple peaks in the spectrum of $^{12}$CO $J=1-0$.
These two sources may consist of two or more velocity components.
Only the main velocity components are fitted since the remaining velocity components
can not be resolved fully in spectra of different species.

The widths of different lines might be the result of the different level of turbulence on different spatial scales.
$\Delta$V of \CCH~$N=1-0$ and \NNHp~$J=1-0$ agrees with each other well (\autoref{kinematics}(c))
because they have similar critical densities \citep{1997ApJ...482..245U}
and both trace the inner dense regions of PGCCs.
The mean width of \CCH~$N=1-0$ is about 1.0 km s$^{-1}$ and it is similar to that of
HCN $J=1-0$  in \cite{2016ApJ...820...37Y} and that of C$^{18}$O $J=1-0$ in \cite{2012ApJ...756...76W}.
It is also close to the mean width of  $^{13}$CO $J=1-0$ for CO-selected cores in the molecular complexes of Orion
(0.9 km s$^{-1}$), Taurus (1.1 km s$^{-1}$), and California (1.4 km s$^{-1}$) \citep{2012ApJS..202....4L,2013ApJS..209...37M}.
For most of these PGCC sources, the line widths seem to be uniform on different scales.
It may indicate that turbulence has  been dissipated on smaller scales. The entire PGCC region is nearly
``transition-to-coherence'' because of the low density, thus cutoff wavelength below which Alfven waves
cannot propagate and support turbulence  \citep{1998ApJ...504..223G} is large.
Below coherence scale, constant residual line widths persist throughout the volume \citep{2005IAUS..231...17T}.
The region outside such a coherent core is more like filled with
cloud components with a radially power-law distributed velocity field.
The Larson's $\Delta$V-r relationship \citep{1981MNRAS.194..809L} can not be applied to PGCCs \citep{2016ApJS..224...43Z}
because the H$_2$ column densities of PGCCs are lowest compared with other star formation samples
such as infrared dark clouds (IRDCs) \citep{2012ApJ...756...76W} thus $^{13}$CO and C$^{18}$O traced
the relatively dense components in PGCCs.
It is also compatible with the concept that these PGCCs are quiescent and most of them seem to be in
transitions from clouds to dense clumps \citep{2012ApJ...756...76W}. The cloud components may contribute to
the broad line widths of C$^{18}$O of several sources with narrow \CCH~lines.
Sources with broad line widths of \CCH~and \NNHp~may be more evolved since sources with emission of \NNHp~stronger
than that of \CCH~all have line widths broader than 1 km s$^{-1}$.

The typical nonthermal velocity traced by \CCH~$N=1-0$ in CO-selected cores
is comparable with that in dense cloud cores \citep{2002ApJ...572..238C},
whose typical line width of NH$_3$ is $\sim$ 0.5 km s$^{-1}$ which corresponds to a $\sigma_{NT}$ $\sim$0.2 km s$^{-1}$.
The ratio between $\sigma_{NT}$ and $\sigma_{therm}$ ranges from 0.7 to 4.7 with a median value of 1.6.
 Among 71 sources, 12 have  $\sigma_{NT}$/$\sigma_{therm}$ $<$ 1.
It is consistent with the idea that supersonic isothermal turbulence is well developed with a lognormal velocity distribution  \citep{2016MNRAS.459.2432V} in outer part of PGCC and decays
as the radius decreases.
The residual line width in coherent region persists due to subsonic or low supersonic turbulence \citep{1998ApJ...504..223G,1998ApJ...496L.109M}.
Emission of \CCH~and \NNHp~traces the inner coherent regions of  PGCCs, where radial density distributions of  pressure-confined
Bonnor-Ebert spheres \citep{2015MNRAS.450.1094P} may be established.

All these characteristics indicate that most of
our selected cores in PGCCs are very cold (with an average gas temperature 14 K),
quiescent and with single component, while still turbulence
dominant. However, there are still some obviously more envolved sources.
Our sample is made up of different components including  clouds,
relatively isolated cold clumps and evolved gas cores.
Emission of \CCH~and \NNHp~originates from the inner dense regions but may have different states of
chemical evolutions (Sect. \ref{abundance_evolution}).

\begin{figure*}
\centering
\includegraphics[width=0.46\linewidth,height=0.35\linewidth]{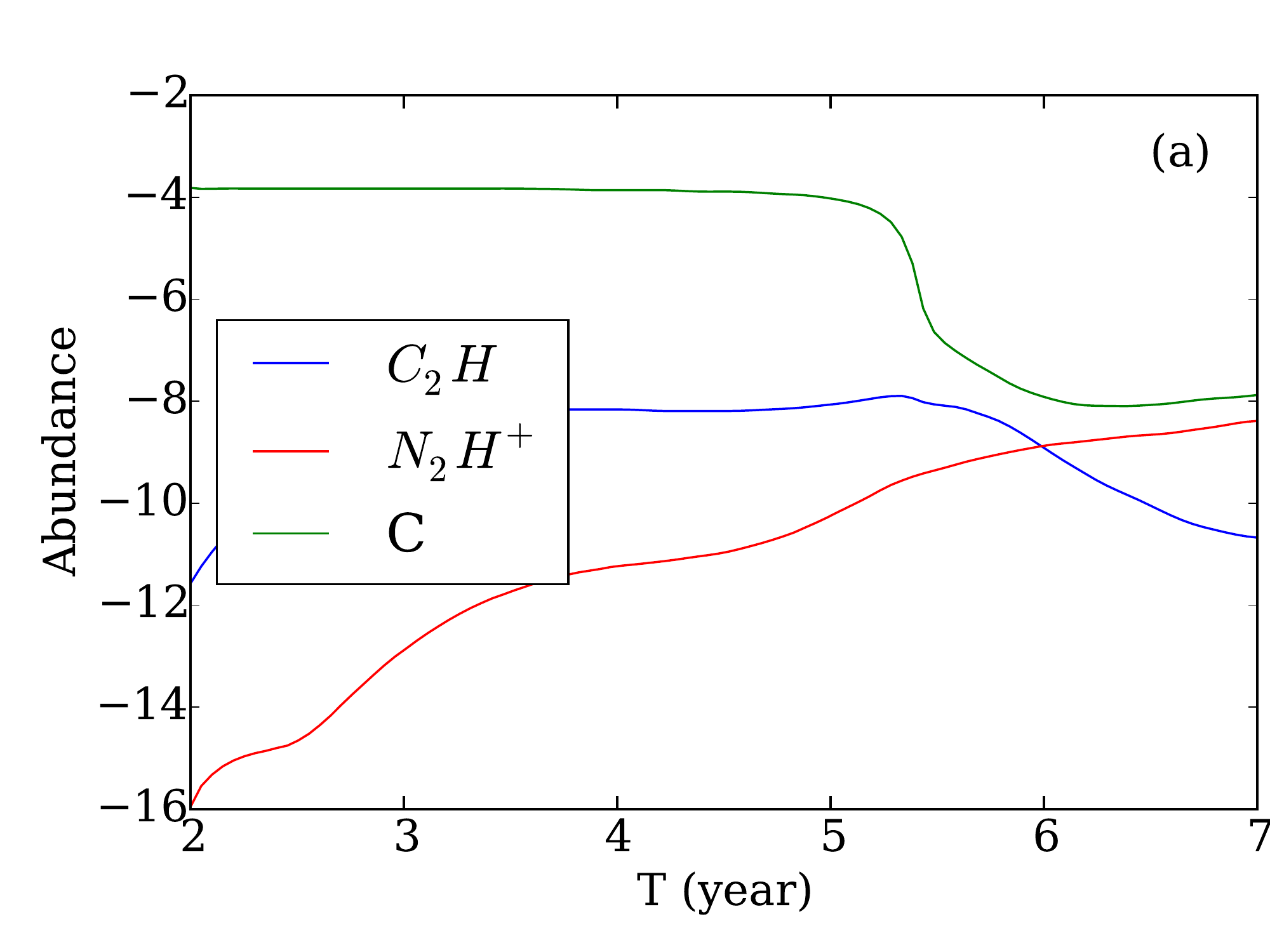}
\includegraphics[width=0.46\linewidth,height=0.35\linewidth]{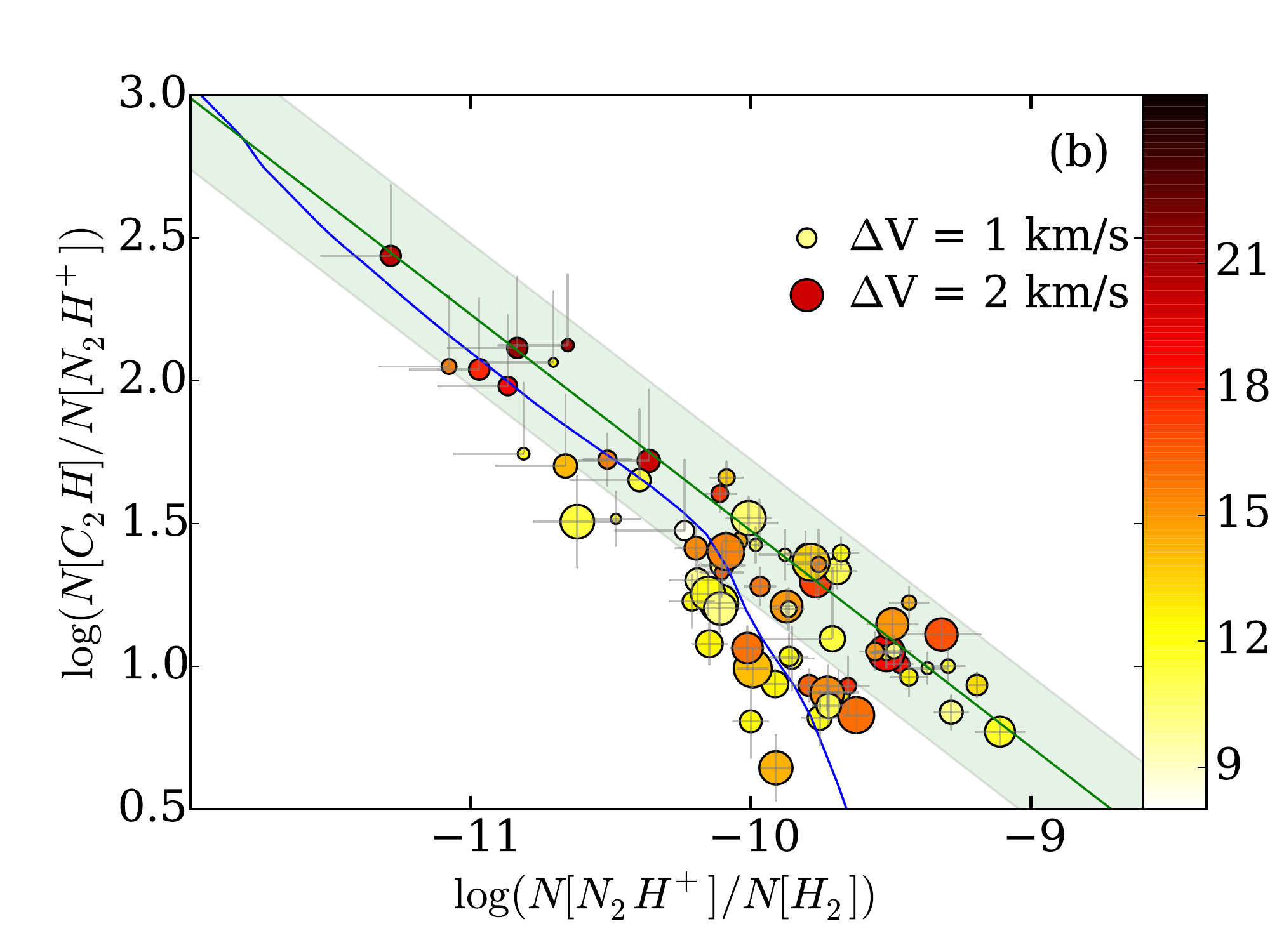}
\caption{\label{fig_mod}
Panel (a): Time evolution of species according to the result of gas-grain chemical model.
Panel (b): Relation between abundances of \NNHp~and $N$[\CCH]/$N$[\NNHp].
The color of each dot represents according CO excitation temperature, and dot-size represents column density
of H$_2$ induced from $^{13}$CO ($N_{CO}$(H$_2$)).
The green line represents the center line of the green band. The blue line shows the result of the gas-grain chemical model.
}
\end{figure*}

\subsection{Abundances of \CCH~and \NNHp} \label{abundance_evolution}
These early cores in our sample with low temperatures but high enough column densities to shield
interstellar radiation field \citep{2017ApJS..228...12T} are good sites to test the evolutions of those two kind of molecules.
\CCH~is generally the most abundant hydrocarbon \citep{2018ApJ...856..151L}
in diffuse molecular gas and dark cloud gas.
It has been known to be a
tracer of photo-dissociation regions (PDRs) \citep{1993A&A...276..473F}.
Recent evidences suggest that it could also trace the cold and dense gas
associated with the early stage of star formation. 
In dark clouds, \CCH~has an extended distribution \citep{2008ApJ...675L..33B}.
In the early stage of dark clouds,  \CCH~is thought to reside in the inner regions
instead of in the external photo-dissociated layers of clumps \citep{2013ApJ...773..123S}.
In the latter stages
\CCH~can still has a high abundance in the outer region when it is oxidized to form other species
such as CO, OH and H$_2$O in the dense center regions \citep{2008ApJ...675L..33B,2014A&A...562A...3M,2016A&A...592A..21F}.
However, \NNHp~usually shows centrally peaked emission for its durability in dense regions.
\NNHp~forms through proton transfer reaction
$N_2+H_3^+\to N_2H^++H_2$ \citep{2001ApJ...552..639A}.
\NNHp~is impeded if CO is present in the gas phase for the competition of CO to react with
its precursor H$_3^+$ through $CO+H_3^+\to HCO^++H_2$.
Furthermore, CO also plays the role of the direct destroyer of \NNHp~through
reaction $N_2H^++CO\to HCO^++N_2$ \citep{2002ApJ...570L.101B}.
Anticorrelation between \NNHp~and
gas-phase CO was presented in envelopes around the pre-stellars and protostars,
such as L1544 \citep{1999ApJ...523L.165C} and IC 5146 \citep{2001ApJ...557..209B}.
In star-forming cores, heating and radiations of protostars will lead to the generation of \CCH~and the destruction of \NNHp, which
make it hard to predict the evolution trends of the abundances of these two species.

\autoref{fig_relation}(a) shows the relation between the abundances  of \CCH~and \NNHp~of CO-selected cores and the
H$_2$ column densities of their host PGCCs ($N^d$(H$_2$)).
The H$_2$ column densities of CO-selected cores derived from emission of
$^{13}$CO $J=1-0$ ($N_{CO}$(H$_2$)) are adopted to calculate the abundances of \CCH~and \NNHp.
$N^d$(H$_2$) are much lower than $N_{CO}$(H$_2$) for relatively large beams ($\sim$4.3\arcmin~at 350 $\mu$m) \citep{2016A&A...594A..28P}
and should be treated as the densities of the environments of CO-selected cores.
\NNHp~abundances of CO-selected cores are positively correlated with $N^d$(H$_2$),
but with a large dispersion (\autoref{fig_relation}(a)).
It suggests that \NNHp~abundances are positively correlated with the evolutionary ages if the cores in PGCCs with larger $N^d$(H$_2$) tend to be more evolved.
On the other hand, the abundances of \CCH~are weakly correlated with  $N^d$(H$_2$).
From \autoref{fig_relation}(a), it is clear to see that
the abundances of \CCH~and \NNHp~of our sources locate in two different regions separated by a green line.
Our sources all have $N$(\CCH) $>$ $N$(\NNHp) thus in pretty young states  ($<$$5\times 10^5$ yr).
The abundances of \CCH~and \NNHp~as well as
their ratios can serve as intrinsic parameters to trace the evolutionary states of PGCC gas cores.
\autoref{fig_relation}(b) shows the number density distribution
of the ratio between the column density of \NNHp~and that
of \CCH~($N$(\NNHp)/$N$(\CCH)). The value of $N$(\NNHp)/$N$(\CCH)
for typical starless cores and IRDCs are also
shown in \autoref{fig_relation}(b). Our PGCC cores generally have
$N$(\NNHp)/$N$(\CCH) higher than those of starless cores such as
TMC-1 and L1498, but lower than those of IRDCs such as G028.34S, which is consistent with the result
of \citep{2017ApJS..228...12T}.

We built a very simple gas-grain chemical model to unveil the evolution  of \CCH~and \NNHp~in cold gas.
In this simulation, a single-point (zero-dimension) chemical code   is run under an ordinary differential equation solver DVODE \citep{doi:10.1137/0910062}
with most physical parameters fixed and dynamical processes are not coupled.
The temperature is adopted as 10 K, the visible extinction $A_v$ = 5, the grain radius $\sigma_g$ = 0.03 $\mu$m, and
the rate of ionization by cosmic-ray $\gamma$ is set as $1.2\times 10^{-17} s^{-1}$ \citep{2004ApJ...617..360L}.
At such a low temperature that is lower than the thermal evaporation temperature of CO 22--25 K \citep{1995ApJ...441..222B,2013MNRAS.431.1296R},
there is nearly no feedback of
gas particles except H$_2$ from grain surfaces, although
grain surface reactions are very active.
The gas phase reactions were downloaded from UMIST Database for Astrochemistry 2012
\citep{2013A&A...550A..36M} with 6\,173 reactions
for 467 kind of species.
The metal abundances are adopted as the low-metal
abundance case of \citet{1982ApJS...48..321G}.
Initially, the elements are all ionized except the hydrogen atoms.
The volume density n(H$_2$) is fixed as 10$^5$ cm$^{-3}$, and the results of the simulation are shown in  \autoref{fig_mod}(a).
Adopting a lower or higher  n(H$_2$) has little influence on the evolution trends of the abundances  except the timescale,
especially for the early stages when the chemical processes are mainly driven by
the atoms and ions
generated from photo-dissociation and photo-ionization during prior more diffuse phase instead of externally induced ionizations.
It is natural that the abundances of species will evolve slower or faster under a lower or higher  n(H$_2$) \citep{2017ApJ...836..194P}.
The values along the x-axis of \autoref{fig_mod}(a) have limited meanings and should not be interpreted as the exact chemical ages considering the variances of volume densities of PGCCs.
Instead, we find that $X$[\CCH]/$X$[\NNHp] is an intrinsic parameter to trace the evolutionary state of a PGCC.

In early stage, the abundance of \NNHp~increases with time while that of \CCH~stays nearly constant.
The abundance of \CCH~drops down quickly after the carbon atoms are depleted, while that of \NNHp~keeps growing.
As shown in \autoref{fig_mod}(b),
the ratio between the abundance of \NNHp~and that of \CCH~can trace the evolution states of PGCCs quite well.
The cores with the lowest abundances of \NNHp~($<$$10^{-10}$) are in the youngest evolutionary states compared with other sources,
and most of them are located in the green band shown in \autoref{fig_mod}(b) whose center line has a power-law index of 0.75.
The power law index is slightly lower than one maybe because of the depletions of CO \citep{2012ApJS..202....4L} and thus the
H$_2$ column densities in evolved regions are underestimated.
For more evolved cores, $X$[\NNHp] versus $X$[\CCH]/$X$[\NNHp] deviates from the
green band for dropping down of the abundances of \CCH. The cores with high abundances of \NNHp~($\sim$$10^{-9}$) but still located in
the green band may have the harshest depletions of CO. Another possibility is that the \CCH~emission regions in these cores are dominated by the
outer regions where the abundances of \CCH~have not yet dropped down.
Observations with higher resolutions will be helpful to
investigate the depletions in the most inner dense regions of PGCCs, and
exam the validity and general applicability of this molecule pair as a tracer of evolutionary state.

\subsection{Emission regions}
Most of our mapped sources are located in nearby star forming regions such as Taurus and Cepheus
\citep[][and the references therein]{2012ApJ...756...76W}.
Cepheus region consists of two velocity components called feature A (300--500 pc) and feature C (800 pc),
and they are  associated with the Gould belt  and the local arm or Orion arm, respectively
\citep{2006MNRAS.369..867O}. The distance of taurus ranges from 130--160 pc \citep{2011RMxAC..40..205L}.
Among these 20 mapped CO-selected cores in PGCCs,
 six  are associated
with IRAS and eight with 2MASS objects, both within CO emission regions. There are no cores associated with both IRAS and 2MASS objects.
This is unlike the case of PGCCs in the second quadrant
(with 98\degr$<$l$<$180\degr~and -4\degr$<$b$<$10\degr~as defined
by \citet{1987ApJ...322..706D}) in which most of the associated IR objects are IRAS point sources
and the rest are 2MASS objects \citep{2016ApJS..224...43Z}.
However, the ratio of the count of cores associated with IR objects to the size
of the sample (referred as core-associated-ratio below) of our sample ($\sim$70\%) is similar to that in
nearby star forming regions \citep{2013ApJS..209...37M}.
However, among the eight cores in Taurus region (see \autoref{tab:map}), only three (G159.2-20A1, G168.7-15A2,  G173.3-16A1)
are associated with IRAS or 2MASS objects. The high core-associated-ratio is mainly contributed by cores
in Cepheus region. Among the eight cores in Cepheus region (see \autoref{tab:map}),
only one (G110.6+09A1) is starless.
The \CCH~and \NNHp~substructures are much smaller than their host clumps and CO emission regions.
The average value of $r$(CO)/$r$(\CCH) and $r$(CO)/$r$(\NNHp)  is 2.3 and 2.9, respectively.
Objects locating within the border line of a substructure are  considered as its associated objects.
In total there are ten substructures associated with IRAS or 2MASS objects. Among them,
six are in Cepheus and only one in Taurus.
The PGCCs in Cepheus region may be generally more evolved than those in the Taurus Complex.

The average value of line widths of substructures in Cepheus, 1.13 km s$^{-1}$, is larger than that in Taurus
region, 0.88 km s$^{-1}$ (\autoref{tab:map}).
The average value of $r$(\CCH)/$r$(\NNHp) for the cores in Taurus, 1.33, is larger than that in Cepheus, 1.15.
Similarly the average value of the $N$(\CCH)/$N$(\NNHp) in Taurus, 23, is larger than that in Cepheus, 13.
These characteristics again confirm that PGCCs in Taurus are less evolved than those in Cepheus.
 For young sources such as PGCCs in Taurus, distribution of \CCH~is much more extended than that of \NNHp. For more evolved sources like PGCCs in Cepheus,
\NNHp~is generated in the dense region \citep{2017ApJS..228...12T} and the area of emission region of \NNHp~continuously expands till close to that of \CCH.

Our finding is compatible with the statistics of \citet{1998ApJ...496L.109M}. Among four complexes, Taurus, Perseus, Orion, and Cepheus,
the percentage of supercritical and cluster associated cores is lowest in Taurus, while highest in Cepheus.
The cores in Cepheus tend to be dynamically and chemically more evolved than those in Taurus, because the Taurus complex is younger than Cepheus complex as a whole.
Another possibility is that the materials in Cepheus are significantly affected  by the large void  between Cassiopeia and
Cepheus \citep{1989ApJ...347..231G,2005A&A...437..919T}.

Among the four mapped cores  not located in Taurus and Cepheus, \CCH~emission regions of two cores (G192.3-11A2, G172.8+02A1) is quite compact, while those of the other two cores (G070.4-01A2, G084.7-01A2) are extended.
Emission of \CCH~with annular distribution is detected in G070.4-01A2 (\autoref{fig_map}).
Similar emission distribution had been detected by \citet{2017ApJS..228...12T} in PGCC with C$_2$S surrounding centrally peaked \NNHp.
Abundance of \CCH~is positively correlated with that of C$_2$S in dark cloud cores beacuse of the reactions
$C_2H^+\stackrel{H_2}{\longrightarrow}C_2H_2{}^+\stackrel{S}{\longrightarrow}C_2S$ \citep{1992ApJ...392..551S}.
Emission of \CCH~shows  annular distribution attributed
to the depletion of \CCH~in the central regions of prestellar cores such as L1498 \citep{2009A&A...505.1199P}.
Similar \CCH~distributions are also observed in various massive star formation regions \citep{2012ApJ...745...47L}
such as  NGC 6334I \citep{2010MNRAS.404.1396W}  and PDRs around \ion{H}{ii}
regions \citep{2013A&A...554A..87P}.
Although the depletion factors of CO are usually not high ($<$ 2) in early PGCC cores \citep{2013ApJ...775L...2L},
the depletion factors
in \CCH~$N=1-0$ emission dominant region can not be ignored considering the dipole moment of \CCH~
(0.77 D; \citealt{1977ApJ...212L..87W}) is about seven times higher
than that of CO.
Depletions in the densest places of PGCCs may produce the annular distributions of \CCH~emission regions.

\section{Summary}
We have made \CCH~$N=1-0$ and \NNHp~$J=1-0$ single-point observations toward
121 CO-selected cores of PGCCs. The detection rate is 59\% for \CCH~$J=1-0$
and 48\% for \NNHp~$J=1-0$. Line parameters were derived through HFS fittings.
Our column densities calculated assuming T$_{ex}$ equal to 5 K and excitation temperatures of CO $J=1-0$  are well consistent with each other,
and most of them (65\%) have deviations of less than 15 percent.
We also mapped  20 sources with the same transitions. Substructures were resolved from maps of \CCH~and \NNHp~with peaks slightly
dislocated. Our main findings are as follows:

1. Most spectra of detected sources are single peaked. Sources that show red and blue profiles
in HCO$^+$ are identified with multicomponents under joint analysis of spectra of \CCH, \NNHp, CO as well as
HCN and HCO$^+$.
Centroid velocities and line widths of \CCH~$N=1-0$ and \NNHp~$J=1-0$ are consistent with each other.
Most sources (83\%) have nonthermal velocities  comparable with or larger than thermal velocities.
All those characteristics indicate that most of
our CO-selected cores in PGCCs are very cold (9--21 K) and quiescent while still dominanted by turbulence.

2. We find that the ratio between the abundance of \CCH~and \NNHp~is a good tracer of evolution for  PGCCs.
Gas grain chemical model based on UMIST network
is applied to fit $N$(\CCH)/$N$(\NNHp) versus $N$(\NNHp).
At the most early stage ($N$(\CCH)/$N$(\NNHp) $>$ 10), abundance of \CCH~is nearly invariable while that  of \NNHp~increases
continuously. Later on ($N$(\CCH)/$N$(\NNHp) $<$ 10), abundance of \NNHp~keeps growing while that of \CCH~drops rapidly as the exhaustion of carbon atoms.
These PGCCs in our sample  are in quite early stages and chemistry driven by
residual atoms
and ions generated from
photo-dissociation and photo-ionization during prior more
diffuse phase still plays a important role.

3. The PGCC cores mapped are approximately virialized ($\alpha$ $<$ 5) and slightly confined by external pressures.
   Sources in Cepheus have lower ratios between $N$(\CCH) and $N$(\NNHp) and larger line widths compared with those in Taurus.
   The probability of finding an associated IR source within PGCC substructures in Cepheus, 55\%, is larger than that in Taurus, 10\%.
   These indicate that PGCCs in Taurus are less chemically evolved than those in Cepheus.
   The \CCH~emission region of G074.4+01A2 shows an annular distribution.

\begin{acknowledgements}
We are grateful to the staff of PMO Qinghai Station.
We also thank Ken'ichi Tatematsu and Junzhi Wang for the helpful discussions.
This project was supported by the grants
	of the National Key R\&D Program of China No. 2017YFA0402600,
    NSFC Nos. 11433008, 11373009, 11373026, 11503035, 11573036 and U1631237, and the China Ministry
	of Science and Technology under State Key Development Program
	for Basic Research (No.2012CB821800), and the
	Top Talents Program of Yunnan Province (2015HA030).
    J. Y. is supported by the Young Researcher Grant of National Astronomical Observatories, Chinese Academy of Sciences.
\end{acknowledgements}

\bibliographystyle{apj}
\bibliography{draft}

\clearpage

\begin{landscape}
\begin{table}
\small
\centering
\caption{Source sample. \label{tab:sources}}
\begin{tabular}{lcccccccccc}
\hline\hline
Designation & RA (J2000) & DEC (J2000)& dist$_r$\footnotemark[1] &dist\footnotemark[2] & $N^d$(H$_2$)
\footnotemark[3]
&$\Delta V(^{13}\mathrm{CO})$&$\Delta V(\mathrm{C^{18}O})$& $\tau(^{13}\mathrm{CO})$ & T$_{ex}$(CO) & $N_{CO}$(H$_2$)\footnotemark[4] \\
       &    &   &kpc  &kpc & $10^{21}\ cm^{-2}$ & km/s&km/s& & K   & $10^{21}\ cm^{-2}$ \\
\hline
G120.1+03A1 & 0:24:23.48 & +65:49:43.4 &1.4(0.3)& 1.2(0.6) &7.62& 2.66(0.02) & 1.99(0.07) & 0.92(0.05) & 15.0(0.7) & 24.2(0.2) \\
G120.6+02A1 & 0:28:55.37 & +65:27:28.6 &--& 0.9(0.3) &4.52& 2.03(0.09) & 1.0(0.2) & 0.7(0.1) & 15(2) & 13.9(0.3) \\
G120.6+02A2 & 0:29:42.75 & +65:26:40.8 &--& 0.9(0.3) &4.52& 2.40(0.04) & 2.0(0.3) & 0.47(0.03) & 17(1) & 14.42(0.01) \\
G130.3+11A1 & 2:32:30.38 & +72:39:22.4 &0.6(0.1)& 0.6(0.2) &6.31& 2.28(0.05) & 1.6(0.2) & 1.0(0.1) & 11.0(0.9) & 11.8(0.2) \\
G133.2+08B2 & 2:51:22.88 & +69:13:10.2 &--& 0.6(0.1) &2.25& 1.7(0.1) & 1.0(0.2) & 0.9(0.2) & 12(3) & 9.9(0.5) \\
G133.2+08B1 & 2:51:45.48 & +69:14:32.8 &--& 0.6(0.1) &2.25& 1.8(0.1) & 1.3(0.4) & 1.4(0.5) & 10(3) & 12(1) \\
G133.4+09A1 & 2:54:32.59 & +69:20:31.8 &--& 0.6(0.1) &9.73& 2.87(0.07) & 2.0(0.2) & 1.1(0.2) & 16(2) & 33(1) \\
G158.8-21A1 & 3:27:38.72 & +30:5:56.4 &0.14(0.01)& 0.14(0.1) &5.06& 1.65(0.07) & 0.7(0.2) & 1.0(0.2) & 15(2) & 16.8(0.8) \\
G158.2-20A1 & 3:30:14.22 & +31:44:56.8 &0.14(0.01)& 0.14(0.1) &0& 1.03(0.03) & 1.0(0.8) & 0.74(0.08) & 18(2) & 11.9(0.3) \\
G159.2-20A1 & 3:33:24.06 & +31:6:59.3 &0.14(0.01)& 0.4(0.1) &13.5& 2.17(0.01) & 1.14(0.02) & 1.5(0.1) & 16.3(0.6) & 38(1) \\
G140.4+06A1 & 3:37:37.02 & +63:7:34.0 &--& 1.2(0.5) &2.45& 2.36(0.07) & 2.1(0.5) & 0.58(0.07) & 14(1) & 10.67(0.04) \\
G159.0-08A1 & 4:9:49.83 & +40:6:55.8 &0.14(0.01)& 0.4(0.1) &1.96& 2.05(0.06) & 1.1(0.1) & 0.70(0.07) & 12(1) & 8.97(0.02) \\
G160.8-09A1 & 4:13:10.09 & +38:10:52.4 &0.14(0.01)& 0.4(0.1) &1.01& 1.0(0.1) & 0.4(0.1) & 0.7(0.1) & 12(3) & 5.2(0.3) \\
G167.2-15A1 & 4:14:29.24 & +29:35:4.5 &0.14(0.01)& 0.2(0.07) &2.66& 1.8(0.2) & 0.6(0.1) & 0.9(0.2) & 10(2) & 7.3(0.3) \\
G168.7-15A2 & 4:18:32.97 & +28:24:41.9 &0.14(0.01)& 0.14(0.1) &6.21& 1.4(0.1) & 0.8(0.1) & 1.4(0.2) & 14(2) & 14(1) \\
G162.4-08A1 & 4:21:40.39 & +37:34:28.1 &0.14(0.01)& 0.14(0.1) &5.29& 2.22(0.08) & 1.3(0.1) & 0.8(0.1) & 10(1) & 8.37(0.01) \\
G150.4+03A2 & 4:24:40.88 & +55:3:2.4 &0.17(0.05)& 0.2(0.1) &0& 1.40(0.05) & 1.2(0.2) & 0.8(0.1) & 12(1) & 7.7(0.1) \\
G173.3-16A1 & 4:29:25.19 & +24:32:45.3 &0.23(0.02)& 0.2(0.1) &6.89& 1.7(0.5) & 0.8(0.1) & 1.3(0.4) & 17(6) & 21(5) \\
G165.6-09A1 & 4:30:57.40 & +34:56:18.8 &0.14(0.01)& 0.5(0.1) &3.02& 1.5(0.2) & 1.0(0.4) & 0.8(0.2) & 13(3) & 7.8(0.6) \\
G174.0-15A1 & 4:32:31.08 & +24:23:25.3 &0.23(0.02)& 0.14(0.1) &8.44& 1.56(0.04) & 0.81(0.10) & 1.1(0.1) & 13(1) & 14.8(0.5) \\
G172.8-14A1 & 4:33:6.18 & +25:58:41.1 &0.23(0.02)& 0.4(0.2) &2.8& 1.77(0.03) & 0.44(0.04) & 1.0(0.1) & 11.9(0.9) & 11.0(0.2) \\
G174.4-15A1 & 4:33:56.07 & +24:10:26.5 &0.23(0.02)& 0.4(0.2) &1.69& 1.5(0.1) & 0.9(0.2) & 1.0(0.1) & 15(2) & 13.5(0.8) \\
G174.7-15A2 & 4:35:36.34 & +24:8:21.2 &0.23(0.02)& 0.2(0.1) &12& 1.15(0.01) & 0.67(0.03) & 1.29(0.07) & 14.1(0.5) & 13.4(0.3) \\
G174.7-15A3 & 4:35:51.79 & +24:8:33.6 &0.23(0.02)& 0.2(0.1) &12& 1.20(0.02) & 0.66(0.09) & 0.84(0.05) & 16.3(0.9) & 11.5(0.2) \\
G173.9-13A1 & 4:39:28.05 & +25:48:13.9 &0.23(0.02)& 0.2(0.1) &6.39& 1.79(0.07) & 1.3(0.1) & 1.0(0.2) & 15(2) & 18.4(0.8) \\
G192.3-11A1 & 5:29:41.06 & +12:21:7.5 &0.40(0.04)& 0.42(0.07) &0.36& 1.63(0.03) & 1.1(0.3) & 0.54(0.03) & 17(1) & 11.46(0.09) \\
G192.3-11A2 & 5:29:54.22 & +12:17:12.5 &0.40(0.04)& 0.42(0.07) &0.36& 1.09(0.02) & 0.92(0.07) & 0.90(0.05) & 18.4(0.9) & 13.9(0.2) \\
G207.3-19A1 & 5:30:42.79 & -4:11:16.0 &0.41(0.05)& 0.60(0.05) &6.8& 1.94(0.02) & 1.1(0.1) & 0.80(0.04) & 20.5(0.9) & 25.1(0.3) \\
G192.2-11A2 & 5:31:28.90 & +12:30:20.8 &0.40(0.04)& 0.42(0.07) &1.48& 1.05(0.06) & 0.5(0.1) & 0.82(0.09) & 20(2) & 15.5(0.7) \\
G192.2-11A3 & 5:31:54.02 & +12:32:36.6 &0.40(0.04)& 0.42(0.04) &1.48& 1.45(0.03) & 1.2(0.1) & 0.74(0.05) & 20(1) & 18.6(0.3) \\
G192.2-11A4 & 5:32:4.48 & +12:31:43.2 &0.40(0.04)& 0.42(0.04) &1.48& 1.22(0.04) & 0.56(0.10) & 0.73(0.07) & 21(2) & 16.1(0.5) \\
G192.1-10A1 & 5:32:34.42 & +12:51:38.8 &0.40(0.04)& 0.42(0.07) &3.26& 1.16(0.04) & 1.1(0.3) & 0.62(0.06) & 21(2) & 14.3(0.3) \\
G172.8+02A1 & 5:36:54.23 & +36:10:33.5 &3.4(0.8)& 1.7(0.1) &5.52& 3.32(0.06) & 2.4(0.3) & 0.65(0.07) & 18(2) & 29.0(0.3) \\
G178.2-00A2 & 5:39:6.75 & +30:4:48.9 &--& 0.96(0.02) &2.72& 2.01(0.04) & 1.4(0.2) & 1.1(0.1) & 11.9(0.8) & 12.5(0.2) \\
G181.8+00A1 & 5:51:30.86 & +27:29:8.1 &--& 1.76(0.04) &3.58& 2.13(0.07) & 0.7(0.2) & 1.0(0.2) & 11(1) & 12.7(0.3) \\
G215.8-17A1 & 5:53:40.87 & -10:23:49.1 &0.41(0.05)& 0.5(0.2) &1.98& 1.52(0.06) & 1.3(0.2) & 1.1(0.2) & 14(1) & 14.4(0.7) \\
G203.2-11A1 & 5:53:44.67 & +3:22:48.9 &0.41(0.05)& 0.42(0.06) &7.61& 1.62(0.04) & 1.27(0.09) & 1.2(0.1) & 12(1) & 15.0(0.6) \\
G202.3-08B2 & 6:0:11.61 & +5:14:52.5 &0.41(0.05)& 0.43(0.06) &5.21& 1.66(0.09) & 0.80(0.07) & 0.47(0.04) & 17(1) & 10.7(0.3) \\
G199.8+00A1 & 6:30:42.73 & +12:1:19.4 &--& 1(1) &7.81& 2.02(0.09) & 1.4(0.1) & 1.1(0.2) & 9(1) & 11.0(0.4) \\
G201.4+00A2 & 6:32:44.05 & +10:29:25.6 &--& 0.42(0.04) &7.47& 5.13(0.09) & 6.4(0.9) & 0.7(0.1) & 12(2) & 21.0(0.2) \\
\hline
\end{tabular}

Only the detected sources are listed.
$^1${Distance adopted from literatures \citep{2012ApJ...756...76W,2016A&A...594A..28P}.}
$^2${Distance given by Bayesian Distance Calculator \citep{2016ApJ...823...77R}.}
$^3${H$_2$ column densities of host PGCCs derived from dust continuum.}
$^4${H$_2$ column densities of CO-selected cores derived from $N$($^{13}$CO).}

\end{table}
\end{landscape}

\addtocounter{table}{-1}

\clearpage
\begin{landscape}
\begin{table}
\small
\centering
\caption{Continued.}
\begin{tabular}{lcccccccccc}
\hline\hline
Designation & RA (J2000) & DEC (J2000)& dist$_r$&dist & $N^d$(H$_2$)
&$\Delta V(^{13}\mathrm{CO})$&$\Delta V(\mathrm{C^{18}O})$& $\tau(^{13}\mathrm{CO})$ & T$_{ex}$(CO) & $N_{CO}$(H$_2$) \\
       &    &     &kpc&kpc & $10^{21}\ cm^{-2}$ & km/s&km/s& & K   & $10^{21}\ cm^{-2}$ \\
\hline
G224.2-00A1 & 7:9:17.49 & -10:28:27.1 &--& 0.42(0.08) &12.6& 3.5(0.1) & 2.2(0.4) & 1.2(0.3) & 12(2) & 27(1) \\
G026.4+08A1 & 18:11:20.92 & -2:0:22.7 &0.8(0.5)& 0.24(0.02) &2.28& 1.01(0.07) & 0.5(0.1) & 1.3(0.3) & 11(2) & 8.8(0.7) \\
G043.0+08A1 & 18:39:21.54 & +12:41:35.3 &0.30(0.07)& 0.4(0.1) &5.77& 0.81(0.05) & 1.5(0.9) & 0.60(0.08) & 11(1) & 2.91(0.03) \\
G038.3-00A1 & 19:4:44.03 & +4:22:51.4 &1.2(0.3)& 2.1(0.2) &7.92& 2.07(0.06) & 1.8(0.2) & 1.0(0.1) & 11(1) & 13.4(0.3) \\
G052.9+03A1 & 19:17:45.44 & +19:15:9.9 &0.7(0.6)& 0.8(0.3) &2.21& 1.67(0.08) & 0.6(0.1) & 1.0(0.1) & 11(1) & 8.7(0.3) \\
G057.1+03A1 & 19:23:55.43 & +23:7:5.3 &--& 0.8(0.3) &5.67& 2.01(0.04) & 1.5(0.1) & 1.1(0.1) & 12.5(0.9) & 15.7(0.3) \\
G058.1+03A1 & 19:26:28.95 & +23:58:37.5 &0.7(0.6)& 0.8(0.3) &6.26& 1.98(0.09) & 0.8(0.2) & 1.2(0.2) & 10(1) & 10.4(0.5) \\
G060.7-01A1 & 19:50:10.65 & +23:55:10.5 &0.9(0.7)& 1.3(0.3) &2.79& 2.18(0.06) & 2.0(0.2) & 2.3(0.6) & 10(1) & 22(3) \\
G070.4-01A2 & 20:14:31.88 & +31:57:5.5 &2.1(0.8)& 2.48(0.08) &7.44& 3.33(0.05) & 2.1(0.3) & 0.81(0.09) & 15(1) & 26.8(0.3) \\
G070.4-01A3 & 20:14:33.09 & +31:57:33.5 &2.1(0.8)& 2.48(0.08) &7.44& 3.32(0.04) & 2.3(0.2) & 0.81(0.08) & 16(1) & 29.4(0.4) \\
G074.1+00A1 & 20:17:56.73 & +35:56:42.8 &3.5(0.8)& 4.3(0.7) &6.42& 3.00(0.07) & 2.5(0.3) & 0.8(0.1) & 14(2) & 29.82(0.06) \\
G102.3+15A2 & 20:35:43.46 & +67:53:18.7 &--& 0.88(0.05) &3.35& 0.91(0.03) & 0.53(0.10) & 1.9(0.3) & 10.5(0.8) & 8.8(0.7) \\
G098.9+13A1 & 20:36:12.70 & +63:52:30.1 &--& 0.7(0.3) &2.06& 1.07(0.05) & 0.6(0.2) & 1.1(0.2) & 9(1) & 5.3(0.2) \\
G093.9+10A2 & 20:37:9.60 & +57:40:28.8 &0.21(0.02)& 0.67(0.02) &4.69& 1.17(0.06) & 1.0(0.2) & 1.9(0.5) & 8(1) & 7.7(0.8) \\
G089.2+04A1 & 20:50:46.66 & +50:26:37.4 &--& 1.64(0.05) &2.01& 3.04(0.09) & 2.2(0.4) & 0.9(0.1) & 10(1) & 11.62(0.07) \\
G084.7-01A3 & 20:56:42.28 & +43:40:42.4 &--& 2.2(0.7) &34.4& 3.3(0.3) & 2.4(0.2) & 1.1(0.3) & 13(3) & 27(1) \\
G084.7-01A2 & 20:56:51.94 & +43:44:15.2 &--& 2.1(0.7) &34.4& 3.34(0.07) & 2.0(0.1) & 1.1(0.2) & 15(2) & 38(2) \\
G111.6+20A1 & 20:57:21.72 & +77:35:12.9 &--& 0.6(0.2) &5.15& 1.31(0.03) & 0.9(0.2) & 1.1(0.1) & 12.6(1.0) & 10.2(0.3) \\
G111.7+20A1 & 20:57:34.44 & +77:35:46.0 &--& 0.6(0.2) &5.15& 1.41(0.06) & 0.9(0.2) & 0.8(0.1) & 15(1) & 10.2(0.2) \\
G103.9+13A1 & 21:2:19.84 & +67:54:0.5 &--& 0.89(0.05) &3.32& 1.31(0.04) & 0.64(0.09) & 1.1(0.1) & 11.8(1.0) & 8.0(0.3) \\
G089.3-00B1 & 21:12:26.16 & +47:24:3.5 &--& 0.67(0.02) &3.26& 3.14(0.07) & 1.4(0.2) & 0.73(0.09) & 11(1) & 11.50(0.03) \\
G105.5+10A1 & 21:41:3.61 & +66:35:17.5 &0.9(0.3)& 0.89(0.07) &2.31& 1.18(0.04) & 0.8(0.3) & 0.67(0.06) & 15(1) & 8.0(0.2) \\
G093.5-04A1 & 21:44:52.90 & +47:37:25.1 &0.12(0.02)& 1(1) &8.17& 2.24(0.06) & 1.6(0.2) & 1.1(0.1) & 12(1) & 15.9(0.4) \\
G093.4-04A2 & 21:46:1.88 & +47:18:6.1 &0.12(0.02)& 1(1) &2.39& 1.89(0.05) & 1.2(0.2) & 0.75(0.08) & 11(1) & 8.66(0.02) \\
G110.6+09A1 & 22:28:17.01 & +69:1:35.1 &0.30(0.03)& 0.8(0.1) &5.39& 1.78(0.03) & 1.02(0.07) & 1.3(0.1) & 12.1(0.7) & 15.2(0.5) \\
G114.1+14A1 & 22:30:13.93 & +75:15:37.6 &--& 0.81(0.08) &7.37& 2.11(0.07) & 2.1(0.5) & 0.63(0.08) & 11(1) & 7.49(0.04) \\
G114.5+14A1 & 22:35:35.60 & +75:17:0.9 &--& 0.81(0.08) &2.48& 1.73(0.08) & 0.9(0.2) & 1.1(0.2) & 12(1) & 12.9(0.6) \\
G114.6+14A1 & 22:38:50.73 & +75:11:45.1 &--& 0.80(0.08) &9.31& 1.65(0.02) & 1.11(0.08) & 1.5(0.2) & 13.2(0.9) & 19.9(0.9) \\
G108.8-00A1 & 22:58:40.27 & +58:56:7.2 &0.18(0.04)& 0.2(0.1) &4.34& 3.21(0.06) & 2.4(0.2) & 0.68(0.07) & 15(1) & 21.5(0.2) \\
G108.8-00A2 & 22:58:54.24 & +58:57:55.3 &0.18(0.04)& 0.2(0.1) &4.34& 2.82(0.04) & 1.9(0.2) & 0.71(0.06) & 15(1) & 22.4(0.2) \\
G115.8-03A1 & 23:56:47.86 & +58:33:36.2 &0.5(0.4)& 0.7(0.1) &1.84& 1.13(0.04) & 0.77(0.09) & 1.1(0.1) & 9.8(0.8) & 5.7(0.1) \\
\hline
\end{tabular}
\end{table}
\end{landscape}

\begin{table*}
\small
\centering
\caption{Line parameters. \label{tab:basic}}
\begin{tabular}{cccccccccccc}
\hline\hline
Designation &          \multicolumn{4}{c}{\CCH}  &            &     \multicolumn{4}{c}{\NNHp} \\
\cline{2-5} \cline{6-10}
            &      $T_a$      &   V$_{\mathrm{LSR}}$   &    $\Delta V$&      $\int T_adV$       &  &  $T_a$         &   V$_{\mathrm{LSR}}$    &    $\Delta V$ &      $\int T_adV$   \\
 &   $\mathrm{K}$ & K km s$^{-1}$ & km s$^{-1}$ & km s$^{-1}$ &    &  $\mathrm{K}$ & K km s$^{-1}$ & km s$^{-1}$ & km s$^{-1}$ \\
\hline
G120.1+03A1&0.24(0.04)&-19.59(0.06)&1.8(0.2)&0.44(0.04)&&0.26(0.04)&-19.70(0.07)&1.5(0.2)&0.59(0.07)\\
G120.6+02A1&0.22(0.02)&-17.39(0.03)&1.05(0.08)&0.23(0.02)&&0.16(0.03)&-17.50(0.05)&0.86(0.09)&0.27(0.04)\\
G120.6+02A2&0.21(0.02)&-17.96(0.05)&1.7(0.1)&0.38(0.02)&&0.21(0.03)&-18.17(0.06)&1.3(0.1)&0.42(0.05)\\
G130.3+11A1&0.42(0.05)&-15.84(0.03)&1.04(0.08)&0.49(0.03)&&0.26(0.05)&-15.93(0.06)&1.1(0.1)&0.50(0.07)\\
G133.2+08B2&0.13(0.02)&-10.50(0.04)&0.76(0.09)&0.10(0.02)&&0.17(0.03)&-10.48(0.06)&0.7(0.1)&0.12(0.02)\\
G133.2+08B1&0.18(0.02)&-10.88(0.04)&1.26(0.10)&0.22(0.02)&&0.12(0.04)&-10.96(0.06)&0.8(0.1)&0.22(0.04)\\
G133.4+09A1&0.68(0.04)&-15.77(0.02)&2.26(0.05)&1.55(0.04)&&0.99(0.04)&-15.69(0.01)&2.07(0.03)&2.5(0.3)\\
G158.8-21A1&0.09(0.03)&5.7(0.1)&1.3(0.3)&0.13(0.02)&&--&--&--&--\\
G158.2-20A1&0.10(0.02)&7.75(0.08)&1.1(0.2)&0.11(0.02)&&--&--&--&--\\
G159.2-20A1&0.50(0.04)&6.83(0.02)&0.66(0.04)&0.41(0.03)&&0.54(0.05)&6.80(0.01)&0.65(0.03)&1.0(0.1)\\
G140.4+06A1&0.06(0.02)&-16.53(0.10)&0.6(0.1)&0.05(0.01)&&0.21(0.05)&-16.68(0.06)&0.6(0.2)&0.25(0.03)\\
G159.0-08A1&0.14(0.05)&-4.41(0.09)&0.7(0.3)&0.13(0.03)&&0.14(0.05)&-4.59(0.04)&0.49(0.08)&0.26(0.05)\\
G160.8-09A1&0.22(0.04)&-4.39(0.04)&0.48(0.09)&0.11(0.02)&&--&--&--&--\\
G167.2-15A1&0.36(0.04)&6.78(0.01)&0.37(0.03)&0.19(0.02)&&0.33(0.04)&6.61(0.04)&0.39(0.09)&0.16(0.02)\\
G168.7-15A2&0.53(0.02)&7.449(0.006)&0.37(0.02)&0.30(0.01)&&0.28(0.05)&7.38(0.02)&0.33(0.06)&0.24(0.03)\\
G162.4-08A1&0.26(0.04)&-1.08(0.04)&1.1(0.1)&0.29(0.03)&&0.46(0.04)&-1.27(0.02)&0.77(0.03)&0.89(0.10)\\
G150.4+03A2&0.15(0.04)&3.39(0.05)&0.3(0.1)&0.05(0.01)&&0.26(0.04)&3.36(0.02)&0.327(0.008)&0.15(0.02)\\
G173.3-16A1&0.73(0.03)&6.424(0.008)&0.66(0.02)&0.49(0.01)&&0.18(0.03)&6.40(0.03)&0.69(0.07)&0.28(0.04)\\
G165.6-09A1&0.45(0.04)&-0.66(0.01)&0.49(0.05)&0.26(0.02)&&0.15(0.04)&-0.75(0.02)&0.43(0.07)&0.24(0.04)\\
G174.0-15A1&0.78(0.03)&5.875(0.006)&0.53(0.02)&0.46(0.01)&&0.17(0.03)&5.83(0.02)&0.44(0.04)&0.23(0.03)\\
G172.8-14A1&0.33(0.04)&5.85(0.02)&0.35(0.01)&0.11(0.02)&&0.15(0.04)&5.78(0.04)&0.3(0.1)&0.07(0.01)\\
G174.4-15A1&0.42(0.03)&7.033(0.009)&0.37(0.02)&0.18(0.01)&&0.11(0.05)&6.95(0.06)&0.3(0.1)&0.07(0.02)\\
G174.7-15A2&1.01(0.04)&5.800(0.007)&0.57(0.02)&0.66(0.02)&&0.53(0.04)&5.797(0.009)&0.53(0.02)&0.9(0.1)\\
G174.7-15A3&0.26(0.02)&6.31(0.02)&0.56(0.04)&0.15(0.01)&&0.14(0.03)&6.28(0.03)&0.44(0.04)&0.16(0.03)\\
G173.9-13A1&0.44(0.03)&6.64(0.01)&0.43(0.03)&0.25(0.02)&&0.12(0.03)&6.53(0.03)&0.52(0.06)&0.21(0.04)\\
G192.3-11A1&0.06(0.02)&12.0(0.2)&1.9(0.4)&0.10(0.02)&&--&--&--&--\\
G192.3-11A2&0.40(0.04)&12.19(0.02)&0.79(0.07)&0.36(0.03)&&0.41(0.05)&12.16(0.02)&0.81(0.06)&0.76(0.08)\\
G207.3-19A1&0.20(0.03)&11.00(0.04)&1.2(0.1)&0.25(0.02)&&--&--&--&--\\
G192.2-11A2&0.50(0.04)&10.35(0.02)&0.59(0.06)&0.36(0.03)&&0.14(0.05)&10.28(0.02)&0.33(0.02)&0.25(0.05)\\
G192.2-11A3&0.23(0.04)&10.46(0.06)&1.2(0.1)&0.29(0.03)&&--&--&--&--\\
G192.2-11A4&0.25(0.04)&10.17(0.05)&1.4(0.1)&0.32(0.03)&&--&--&--&--\\
G192.1-10A1&0.11(0.02)&10.30(0.07)&1.5(0.2)&0.18(0.02)&&--&--&--&--\\
G172.8+02A1&0.33(0.05)&-17.48(0.05)&1.9(0.1)&0.72(0.03)&&0.57(0.05)&-17.44(0.04)&1.9(0.1)&1.4(0.2)\\
G178.2-00A2&0.15(0.04)&-0.80(0.06)&0.7(0.1)&0.14(0.03)&&0.22(0.05)&-0.84(0.04)&0.7(0.1)&0.45(0.06)\\
G181.8+00A1&0.37(0.04)&2.60(0.03)&1.10(0.09)&0.40(0.03)&&0.38(0.05)&2.48(0.03)&0.85(0.07)&0.78(0.08)\\
G215.8-17A1&0.16(0.03)&9.38(0.03)&0.7(0.1)&0.14(0.02)&&--&--&--&--\\
G203.2-11A1&0.20(0.05)&10.55(0.06)&1.1(0.2)&0.23(0.03)&&0.32(0.05)&10.54(0.05)&0.9(0.1)&0.60(0.09)\\
G202.3-08B2&0.18(0.04)&12.13(0.04)&0.52(0.10)&0.15(0.03)&&0.17(0.05)&12.00(0.03)&0.45(0.06)&0.40(0.08)\\
G199.8+00A1&0.19(0.03)&6.24(0.03)&0.55(0.06)&0.14(0.02)&&0.12(0.03)&6.10(0.09)&0.9(0.1)&0.15(0.03)\\
G201.4+00A2&0.11(0.04)&4.92(0.05)&0.6(0.1)&0.25(0.04)&&0.19(0.05)&4.67(0.04)&0.71(0.09)&0.33(0.06)\\
\hline
\end{tabular}
\end{table*}

\addtocounter{table}{-1}

\begin{table*}
\small
\centering
\caption{Continued.}
\begin{tabular}{cccccccccccc}
\hline\hline
Designation &          \multicolumn{4}{c}{\CCH}  &            &     \multicolumn{4}{c}{\NNHp} \\
\cline{2-5} \cline{6-10}
            &      $T_a$      &   V$_{\mathrm{LSR}}$   &    $\Delta V$&      $\int T_adV$       &  &  $T_a$         &   V$_{\mathrm{LSR}}$    &    $\Delta V$ &      $\int T_adV$   \\
 &   $\mathrm{K}$ & K km s$^{-1}$ & km s$^{-1}$ & km s$^{-1}$ &    &  $\mathrm{K}$ & K km s$^{-1}$ & km s$^{-1}$ & km s$^{-1}$ \\
\hline
G224.2-00A1&0.15(0.03)&14.04(0.05)&1.4(0.1)&0.32(0.02)&&0.17(0.05)&13.83(0.10)&1.4(0.3)&0.39(0.07)\\
G026.4+08A1&0.15(0.03)&6.37(0.05)&0.6(0.1)&0.07(0.01)&&--&--&--&--\\
G043.0+08A1&0.17(0.03)&4.09(0.02)&0.35(0.03)&0.07(0.01)&&--&--&--&--\\
G038.3-00A1&0.42(0.04)&16.97(0.03)&1.24(0.09)&0.56(0.03)&&0.85(0.05)&16.80(0.02)&1.33(0.06)&1.9(0.2)\\
G052.9+03A1&0.22(0.04)&10.76(0.04)&0.55(0.08)&0.13(0.02)&&0.22(0.05)&10.79(0.04)&0.70(0.10)&0.34(0.06)\\
G057.1+03A1&0.12(0.02)&12.22(0.04)&1.2(0.1)&0.15(0.01)&&0.19(0.01)&12.24(0.02)&1.00(0.05)&0.38(0.04)\\
G058.1+03A1&0.51(0.04)&10.18(0.02)&0.55(0.04)&0.36(0.03)&&0.40(0.05)&10.09(0.02)&0.54(0.02)&0.70(0.08)\\
G060.7-01A1&0.34(0.05)&11.11(0.03)&0.69(0.07)&0.27(0.03)&&0.17(0.05)&11.18(0.05)&0.62(0.10)&0.37(0.07)\\
G070.4-01A2&0.15(0.04)&11.24(0.09)&2.1(0.2)&0.34(0.04)&&0.40(0.03)&10.70(0.03)&1.71(0.09)&0.89(0.10)\\
G070.4-01A3&0.14(0.05)&11.2(0.1)&2.1(0.3)&0.03(0.05)&&0.54(0.06)&10.77(0.04)&1.5(0.1)&1.2(0.1)\\
G074.1+00A1&0.11(0.03)&-1.3(0.1)&2.3(0.3)&0.26(0.03)&&0.22(0.03)&-0.73(0.06)&2.1(0.2)&0.57(0.07)\\
G102.3+15A2&0.56(0.02)&2.756(0.009)&0.61(0.02)&0.34(0.01)&&0.44(0.03)&2.641(0.006)&0.47(0.01)&0.8(0.1)\\
G098.9+13A1&0.33(0.03)&-2.67(0.02)&0.50(0.06)&0.17(0.02)&&0.11(0.03)&-2.81(0.03)&0.37(0.05)&0.15(0.03)\\
G093.9+10A2&0.25(0.03)&-2.06(0.02)&0.49(0.05)&0.13(0.01)&&--&--&--&--\\
G089.2+04A1&0.30(0.04)&-4.41(0.03)&1.11(0.09)&0.36(0.03)&&0.12(0.03)&-4.81(0.09)&1.0(0.2)&0.24(0.04)\\
G084.7-01A3&0.44(0.04)&0.69(0.02)&1.59(0.05)&0.91(0.04)&&0.36(0.03)&0.88(0.03)&1.4(0.1)&0.9(0.1)\\
G084.7-01A2&0.61(0.04)&0.96(0.02)&1.78(0.04)&1.38(0.04)&&0.83(0.04)&0.70(0.01)&1.70(0.05)&2.1(0.3)\\
G111.6+20A1&0.33(0.05)&-8.02(0.03)&0.78(0.10)&0.31(0.03)&&0.38(0.05)&-8.14(0.02)&0.70(0.04)&0.72(0.10)\\
G111.7+20A1&0.32(0.04)&-8.06(0.03)&0.72(0.07)&0.26(0.03)&&0.29(0.04)&-8.15(0.02)&0.57(0.04)&0.51(0.06)\\
G103.9+13A1&0.46(0.05)&2.93(0.02)&0.59(0.05)&0.37(0.03)&&0.47(0.05)&2.99(0.02)&0.67(0.06)&0.79(0.09)\\
G089.3-00B1&0.18(0.03)&2.26(0.04)&1.0(0.1)&0.15(0.02)&&0.22(0.03)&2.24(0.03)&0.89(0.07)&0.45(0.05)\\
G105.5+10A1&0.29(0.05)&-10.21(0.04)&0.7(0.1)&0.26(0.04)&&0.13(0.05)&-10.24(0.03)&0.49(0.07)&0.25(0.07)\\
G093.5-04A1&0.18(0.03)&3.62(0.04)&0.73(0.08)&0.12(0.01)&&0.13(0.03)&3.53(0.03)&0.56(0.08)&0.23(0.03)\\
G093.4-04A2&0.27(0.03)&5.32(0.02)&0.43(0.04)&0.15(0.01)&&--&--&--&--\\
G110.6+09A1&0.25(0.04)&-4.43(0.05)&0.9(0.1)&0.21(0.03)&&0.24(0.04)&-4.47(0.05)&0.92(0.09)&0.42(0.06)\\
G114.1+14A1&0.15(0.03)&-4.01(0.05)&0.5(0.1)&0.06(0.01)&&0.06(0.03)&-3.56(0.08)&0.42(0.06)&0.04(0.01)\\
G114.5+14A1&0.72(0.04)&-4.94(0.01)&0.77(0.04)&0.59(0.03)&&0.34(0.04)&-4.98(0.03)&0.72(0.06)&0.54(0.07)\\
G114.6+14A1&0.71(0.04)&-3.66(0.02)&1.21(0.04)&0.93(0.03)&&1.11(0.04)&-3.64(0.01)&1.26(0.03)&2.2(0.2)\\
G108.8-00A1&0.17(0.03)&-50.38(0.05)&1.9(0.2)&0.37(0.03)&&0.16(0.04)&-50.61(0.09)&1.3(0.2)&0.32(0.05)\\
G108.8-00A2&0.12(0.03)&-49.20(0.07)&1.7(0.2)&0.21(0.02)&&0.19(0.04)&-49.39(0.07)&1.4(0.2)&0.39(0.06)\\
G115.8-03A1&0.24(0.04)&-0.64(0.02)&0.35(0.02)&0.12(0.02)&&0.31(0.04)&-0.74(0.04)&0.45(0.08)&0.16(0.02) \\
\hline
\end{tabular}
\end{table*}

\clearpage

\begin{table*}
\small
\centering
\caption{Derived parameters. \label{tab:derived}}
\begin{tabular}{cccccccccc}
\hline\hline
Designation    &   $\sigma_{NT}(C_2H)$   &$\sigma_{NT}(N_2H^+)$ & \multicolumn{3}{c}{$T_{ex}=T_{ex}$(CO)}   &         &     \multicolumn{3}{c}{$T_{ex}$=5 K} \\
\cline{4-6} \cline{8-10}
                     &  &               & $N$(\CCH)    &  $N$(\NNHp)        &   ratio     &        & $N$(\CCH)    &  $N$(\NNHp)        & ratio  \\
            & km s$^{-1}$ &  km s$^{-1}$  & $10^{12}\ cm^{-2}$ & $10^{11}\ cm^{-2}$&         &        &  $10^{12}\ cm^{-2}$ & $10^{11}\ cm^{-2}$&     \\
\hline
G120.1+03A1&0.76(0.07)&0.6(0.1)&52(5)&32(3)&16(2)&&53(5)&34(4)&15(2)\\
G120.6+02A1&0.44(0.04)&0.36(0.04)&28(2)&15(2)&19(3)&&28(2)&14(2)&19(3)\\
G120.6+02A2&0.71(0.06)&0.54(0.06)&48(3)&24(3)&19(2)&&45(2)&23(2)&19(2)\\
G130.3+11A1&0.44(0.04)&0.45(0.06)&52(3)&24(3)&21(3)&&66(4)&29(4)&22(3)\\
G133.2+08B2&0.32(0.04)&0.27(0.05)&10(1)&6(1)&16(4)&&10(1)&6(1)&16(4)\\
G133.2+08B1&0.53(0.04)&0.34(0.05)&23(1)&10(1)&22(4)&&26(1)&11(2)&22(4)\\
G133.4+09A1&0.96(0.02)&0.88(0.01)&204(5)&158(19)&12(1)&&271(7)&507(61)&5(1)\\
G158.8-21A1&0.6(0.1)&--&15(3)&(1)&--&&14(2)&(1)&--\\
G158.2-20A1&0.46(0.07)&--&15(2)&(2)&--&&12(1)&(1)&--\\
G159.2-20A1&0.27(0.02)&0.27(0.01)&53(3)&61(7)&8(1)&&60(4)&76(9)&8(1)\\
G140.4+06A1&0.26(0.05)&0.23(0.07)&5(1)&13(1)&4(1)&&5(1)&13(1)&4(1)\\
G159.0-08A1&0.3(0.1)&0.20(0.03)&13(3)&12(2)&10(3)&&14(3)&13(2)&10(3)\\
G160.8-09A1&0.19(0.04)&--&11(2)&(1.0)&--&&13(2)&(1)&--\\
G167.2-15A1&0.15(0.01)&0.16(0.04)&20(2)&7.6(0.8)&26(4)&&25(2)&9(1)&26(4)\\
G168.7-15A2&0.14(0.01)&0.12(0.02)&36(1)&13(1)&27(4)&&45(2)&14(2)&31(4)\\
G162.4-08A1&0.45(0.05)&0.32(0.01)&30(3)&43(4)&6(1)&&36(3)&60(6)&6(1)\\
G150.4+03A2&0.14(0.05)&0.13(0.01)&4(1)&7(1)&6(2)&&5(1)&8(1)&5(2)\\
G173.3-16A1&0.27(0.01)&0.28(0.03)&66(1)&16(2)&40(6)&&92(2)&15(2)&59(6)\\
G165.6-09A1&0.20(0.02)&0.17(0.03)&30(2)&12(2)&24(4)&&36(2)&12(2)&28(4)\\
G174.0-15A1&0.22(0.01)&0.17(0.01)&55(1)&12(1)&45(6)&&92(2)&12(1)&71(6)\\
G172.8-14A1&0.14(0.01)&0.13(0.05)&11(1)&3.6(0.7)&32(8)&&14(2)&4.0(0.8)&34(8)\\
G174.4-15A1&0.14(0.01)&0.12(0.06)&22(1)&4.2(0.9)&53(13)&&24(1)&4.0(0.9)&61(13)\\
G174.7-15A2&0.23(0.01)&0.22(0.01)&82(2)&49(6)&16(2)&&130(4)&64(8)&20(2)\\
G174.7-15A3&0.23(0.02)&0.17(0.02)&19(1)&9(1)&21(4)&&18(1)&8(1)&21(4)\\
G173.9-13A1&0.17(0.01)&0.21(0.03)&30(2)&11(2)&25(5)&&34(2)&11(2)&30(5)\\
G192.3-11A1&0.8(0.2)&--&13(3)&(1)&--&&11(2)&(1)&--\\
G192.3-11A2&0.33(0.03)&0.34(0.02)&48(3)&47(5)&10(1)&&48(3)&49(5)&9(1)\\
G207.3-19A1&0.50(0.05)&--&35(2)&(1)&--&&29(2)&(1)&--\\
G192.2-11A2&0.24(0.02)&0.12(0.01)&52(4)&16(3)&31(6)&&51(4)&13(2)&38(6)\\
G192.2-11A3&0.50(0.06)&--&42(4)&(8)&--&&35(3)&(6)&--\\
G192.2-11A4&0.58(0.06)&--&47(4)&(3)&--&&39(4)&(2)&--\\
G192.1-10A1&0.65(0.07)&--&27(2)&(2)&--&&20(1)&(1)&--\\
G172.8+02A1&0.79(0.05)&0.81(0.05)&98(4)&89(11)&11(1)&&93(4)&104(13)&9(1)\\
G178.2-00A2&0.31(0.05)&0.28(0.05)&14(3)&22(3)&6(1)&&15(3)&25(3)&6(1)\\
G181.8+00A1&0.46(0.04)&0.36(0.03)&43(3)&38(4)&11(1)&&53(3)&49(5)&10(1)\\
G215.8-17A1&0.30(0.05)&--&15(2)&(3)&--&&16(2)&(3)&--\\
G203.2-11A1&0.46(0.07)&0.39(0.05)&25(3)&30(4)&8(1)&&27(3)&36(5)&7(1)\\
G202.3-08B2&0.21(0.04)&0.18(0.03)&20(3)&23(4)&8(2)&&18(3)&21(4)&8(2)\\
G199.8+00A1&0.23(0.03)&0.39(0.06)&14(1)&7(1)&20(5)&&16(2)&8(1)&20(5)\\
G201.4+00A2&0.27(0.05)&0.29(0.04)&26(4)&16(2)&16(3)&&28(4)&18(3)&15(3)\\
\hline
\end{tabular}
\end{table*}

\addtocounter{table}{-1}

\begin{table*}
\small
\centering
\caption{Continued.}
\begin{tabular}{cccccccccc}
\hline\hline
Designation    &   $\sigma_{NT}(C_2H)$   &$\sigma_{NT}(N_2H^+)$ & \multicolumn{3}{c}{$T_{ex}=T_{ex}$(CO)}   &         &     \multicolumn{3}{c}{$T_{ex}$=5 K} \\
\cline{4-6} \cline{8-10}
                     &  &               & $N$(\CCH)    &  $N$(\NNHp)        &   ratio     &        & $N$(\CCH)    &  $N$(\NNHp)        & ratio  \\
            & km s$^{-1}$ &  km s$^{-1}$  & $10^{12}\ cm^{-2}$ & $10^{11}\ cm^{-2}$&         &        &  $10^{12}\ cm^{-2}$ & $10^{11}\ cm^{-2}$&     \\
\hline
G224.2-00A1&0.61(0.04)&0.6(0.1)&34(2)&19(3)&17(3)&&36(2)&21(3)&17(3)\\
G026.4+08A1&0.24(0.05)&--&7(1)&(1)&--&&8(1)&(1)&--\\
G043.0+08A1&0.14(0.01)&--&7(1)&(5)&--&&7(1)&(6)&--\\
G038.3-00A1&0.53(0.04)&0.56(0.03)&61(3)&104(11)&5.9(0.7)&&77(4)&217(23)&3.6(0.7)\\
G052.9+03A1&0.23(0.03)&0.29(0.04)&13(2)&16(2)&8(2)&&15(2)&19(3)&7(2)\\
G057.1+03A1&0.51(0.05)&0.42(0.02)&16(1)&19(1)&8(1)&&17(1)&21(2)&8(1)\\
G058.1+03A1&0.23(0.02)&0.22(0.01)&38(2)&33(3)&11(1)&&53(3)&44(5)&11(1)\\
G060.7-01A1&0.29(0.03)&0.26(0.04)&27(3)&17(3)&16(3)&&34(3)&20(3)&16(3)\\
G070.4-01A2&0.90(0.09)&0.72(0.04)&40(4)&50(5)&8(1)&&38(4)&57(6)&6(1)\\
G070.4-01A3&0.9(0.1)&0.62(0.05)&47(5)&69(6)&6(1)&&43(5)&87(8)&5(1)\\
G074.1+00A1&1.0(0.1)&0.91(0.07)&29(3)&30(3)&9(1)&&29(3)&32(4)&9(1)\\
G102.3+15A2&0.25(0.01)&0.19(0.01)&36(1)&37(4)&9(1)&&52(1)&50(6)&10(1)\\
G098.9+13A1&0.21(0.02)&0.15(0.02)&17(1)&7(1)&24(5)&&21(2)&8(1)&26(5)\\
G093.9+10A2&0.20(0.02)&--&13(1)&(4)&--&&16(1)&(5)&--\\
G089.2+04A1&0.47(0.04)&0.43(0.09)&37(2)&11(2)&33(6)&&45(3)&13(2)&34(6)\\
G084.7-01A3&0.67(0.02)&0.60(0.04)&105(4)&45(6)&23(3)&&126(4)&53(7)&23(3)\\
G084.7-01A2&0.75(0.02)&0.72(0.02)&170(4)&121(15)&14(1)&&224(5)&225(28)&10(1)\\
G111.6+20A1&0.33(0.04)&0.29(0.02)&34(3)&37(5)&9(1)&&39(4)&45(6)&8(1)\\
G111.7+20A1&0.30(0.03)&0.23(0.02)&31(3)&28(3)&11(1)&&33(3)&30(3)&11(1)\\
G103.9+13A1&0.24(0.02)&0.28(0.03)&40(3)&40(4)&10(1)&&52(4)&53(6)&9(1)\\
G089.3-00B1&0.40(0.06)&0.37(0.03)&15(1)&21(2)&7(1)&&18(2)&25(3)&7(1)\\
G105.5+10A1&0.31(0.05)&0.20(0.03)&31(4)&14(4)&22(7)&&32(4)&13(4)&23(7)\\
G093.5-04A1&0.30(0.04)&0.23(0.03)&13(1)&11(1)&12(2)&&14(1)&12(1)&11(2)\\
G093.4-04A2&0.17(0.02)&--&15(1)&(3)&--&&18(1)&(3)&--\\
G110.6+09A1&0.36(0.05)&0.39(0.04)&22(3)&20(3)&10(2)&&25(3)&24(3)&10(2)\\
G114.1+14A1&0.20(0.06)&0.17(0.03)&5(1)&1.8(0.7)&32(15)&&6(1)&2.0(0.8)&32(15)\\
G114.5+14A1&0.32(0.02)&0.30(0.02)&67(2)&27(3)&24(3)&&108(4)&33(4)&32(3)\\
G114.6+14A1&0.51(0.02)&0.53(0.01)&109(3)&127(13)&8.6(0.9)&&169(5)&119(12)&14.3(0.9)\\
G108.8-00A1&0.82(0.07)&0.56(0.10)&44(3)&17(3)&25(4)&&42(2)&17(3)&24(4)\\
G108.8-00A2&0.71(0.10)&0.59(0.08)&25(3)&21(3)&11(2)&&23(2)&21(3)&10(2)\\
G115.8-03A1&0.14(0.01)&0.18(0.04)&12(1)&7.8(0.8)&15(3)&&14(2)&10(1)&14(3)\\
\hline
\end{tabular}
\end{table*}

\clearpage
\begin{landscape}
\begin{table}
\small
\centering
\caption{Mapping parameters. \label{tab:map}}
\begin{tabular}{ccccccccccccccccc}
\hline\hline
              &    & & \multicolumn{5}{c}{\CCH}    &         &  \multicolumn{5}{c}{\NNHp}                            &               &            &          \\
\cline{4-8} \cline{10-14}
Designation\footnotemark[1]  &sub&$r(CO)$& center  & V$_{\mathrm{LSR}}$ &    $\Delta$V  &   $r$ &    N && center  & V$_{\mathrm{LSR}}$ &    $\Delta$V  &   $r$ &    N &   $M_{sub}$\footnotemark[2] &   $M_{vir}$&  $\alpha$\\
              &    &\arcmin& (\arcmin,\arcmin)   & km s$^{-1}$ &  km s$^{-1}$  &  \arcmin &    cm$^{-2}$ && (\arcmin,\arcmin)  & km s$^{-1}$ &  km s$^{-1}$  &   \arcmin &   cm$^{-2}$ & M${_\sun}$ &  $M_{\sun}$  & \\
\hline
G130.3+11A1$^\Delta$ & -- &2.5&(-0.5,-0.5)& -15.55 & 1.55 & 1.35 & 64(5)&  &(-0.0,-0.7)& -15.81 & 1.48 & 1.07 & 3.5(0.3)& 7.3(0.7)& 57(5)& 7.9 \\
G133.4+09A1$^\Delta$ & -- &2.7&(0.1,0.6)& -16.17 & 2.95 & 1.55 & 236(21)&  &(-0.2,-0.1)& -15.58 & 2.7 & 1.47 & 11(1)& 29(2)& 285(26)& 9.6 \\
G159.2-20A1$^\star$ & -- &3.2&(-0.6,0.6)& 6.81 & 1.23 & 1.8 & 89(8)&  &(-0.9,0.4)& 6.73 & 2.03 & 1.42 & 14(1)& 6.3(0.6)& 29(2)& 4.6 \\
G167.2-15A1$^\star$ & -- &2.4&(-0.6,0.5)& 6.54 & 1.01 & 1.04 & 40(3)&  &(-0.5,0.5)& 7.45 & 0.67 & 0.49 & 0.51(0.05)& 0.56(0.05)& 5.5(0.5)& 9.7 \\
G168.7-15A2$^\star$ & -- &--&(0.4,2.0)& 7.28 & 0.62 & 1.96 & 48(4)&  &(0.0,2.5)& 7.46 & 2.03 & 1.3 & 5.9(0.5)& 0.62(0.06)& 2.5(0.3)& 4.0 \\
G173.3-16A1$^\star$ & -- &--&(-0.8,0.3)& 6.31 & 0.66 & 1.3 & 80(7)&  &(-0.9,0.1)& 6.14 & 2.44 & 0.81 & 3.4(0.3)& 0.54(0.05)& 1.8(0.2)& 3.3 \\
G165.6-09A1$^\star$ & -- &--&(-0.6,-0.3)& -0.74 & 1.24 & 0.91 & 39(3)&  &(-0.9,-0.5)& -0.47 & 1.63 & 0.74 & 3.8(0.3)& 0.88(0.08)& 19(1)& 21.8 \\
G174.0-15A1$^\star$ & W &4.5&(1.0,0.6)& 6.04 & 0.73 & 1.44 & 80(7)&  &(1.2,0.6)& 5.88 & 1.36 & 0.62 & 2.1(0.2)& 0.55(0.05)& 1.8(0.2)& 3.3 \\
G174.0-15A1$^\star$ & E &4.5&(4.1,1.0)& 6.14 & 0.85 & 2.07 & 86(7)&  &(3.5,1.9)& 6.24 & 1.84 & 1.22 & 6.4(0.6)& 1.1(0.1)& 4.5(0.4)& 4.1 \\
G174.7-15A2$^\star$ & -- &3.5&(0.1,1.3)& 5.95 & 0.63 & 1.47 & 99(8)&  &(-0.4,1.0)& 5.97 & 1.73 & 1.25 & 5.9(0.5)& 0.86(0.08)& 3.7(0.2)& 4.3 \\
G173.9-13A1$^\star$ & S &3.0&(2.1,0.1)& 6.55 & 0.82 & 1.53 & 57(5)&  &(2.5,-1.6)& 6.46 & 1.69 & 0.39 & 1.7(0.2)& 0.41(0.04)& 1.5(0.2)& 3.6\\
G173.9-13A1$^\star$ & N &2.5&(-3.0,3.6)& 6.19 & 0.74 & 1.39 & 56(5)&  &(-2.1,3.1)& 6.6 & 1.83 & 0.89 & 4.2(0.4)& 0.40(0.04)& 2.8(0.3)& 7.0 \\
G192.3-11A2 & -- &1.7&(0.4,-0.8)& 11.73 & 0.7 & 0.87 & 95(8)&  &(0.0,-0.4)& 11.98 & 0.7 & 0.63 & 10.9(1.0)& 1.5(0.1)& 4.7(0.4)& 3.3 \\
G172.8+02A1 & -- &2.3&(-0.4,0.0)& -17.27 & 2.28 & 0.89 & 91(8)&  &(-0.4,0.3)& -17.38 & 2.43 & 0.9 & 6.9(0.6)& 29(2)& 294(26)& 10.1 \\
G070.4-01A2 & -- &3.6&(2.2,-0.1)& 11.24 & 1.8 & 1.66 & 35(3)&  &(0.4,0.2)& 11.24 & 2.0 & 0.89 & 4.1(0.4)& 90(8)& 262(24)& 2.9 \\
G102.3+15A2$^\Delta$ & SW &3.0&(0.1,-0.1)& 2.66 & 0.49 & 0.8 & 42(3)&  &(-0.0,-0.4)& 2.93 & 2.27 & 0.78 & 4.6(0.4)& 2.7(0.2)& 6.0(0.5)& 2.2 \\
G102.3+15A2$^\Delta$ & NE &3.0&(1.5,1.1)& 2.47 & 0.8 & 0.5 & 33(3)&  &(1.3,0.8)& 2.32 & 1.35 & 0.51 & 1.8(0.2)& 0.83(0.07)& 10.5(0.9)& 12.7 \\
G084.7-01A2 & NW &3.5&(-0.3,-0.7)& 1.09 & 2.27 & 2.39 & 150(13)&  &(-0.9,-1.2)& 1.41 & 2.36 & 2.1 & 11(1)& 695(63)& 851(77)& 1.2 \\
G084.7-01A2 & SE &3.5&(1.9,-3.2)& 0.6 & 2.85 & 1.2 & 151(14)&  &(2.5,-2.9)& 3.38 & 8.3 & 1.16 & 20(1)& 156(14)& 741(67)& 4.7 \\
G111.6+20A1$^\Delta$ & -- &3.6&(-0.1,0.9)& -7.92 & 0.86 & 1.27 & 68(6)&  &(-0.4,0.7)& -7.81 & 1.95 & 0.96 & 9.3(0.8)& 7.0(0.6)& 16(1)& 2.3 \\
G103.9+13A1$^\Delta$ & -- &2.7&(0.5,0.8)& 2.84 & 1.27 & 1.32 & 54(4)&  &(0.2,0.5)& 3.26 & 1.98 & 0.99 & 4.7(0.4)& 10.4(0.9)& 52(4)& 5.0 \\
G110.6+09A1$^\Delta$ & -- &2.0&(-0.9,-0.5)& -4.25 & 0.84 & 0.83 & 65(5)&  &(-1.0,-0.8)& -4.18 & 2.32 & 0.95 & 8.1(0.7)& 4.8(0.4)& 20(1)& 4.2 \\
G114.5+14A1$^\Delta$ & W &2.9&(-0.4,-0.0)& -4.79 & 1.0 & 0.85 & 92(8)&  &(-0.5,-0.1)& -4.48 & 2.29 & 0.58 & 4.3(0.4)& 6.6(0.6)& 17(1)& 2.6 \\
G114.5+14A1$^\Delta$ & E &2.9&(0.5,0.4)& -4.89 & 0.82 & 0.68 & 58(5)&  &(0.5,0.1)& -4.68 & 1.58 & 0.51 & 3.4(0.3)& 3.4(0.3)& 10.2(0.9)& 3.0 \\
G114.6+14A1$^\Delta$ & W &3.2&(0.1,-0.2)& -3.71 & 1.01 & 0.91 & 80(7)&  &(-0.2,-0.3)& -3.77 & 1.89 & 0.94 & 9.5(0.9)& 8.3(0.7)& 28(2)& 3.4 \\
G114.6+14A1$^\Delta$ & E &3.2&(1.2,0.0)& -3.42 & 0.82 & 0.98 & 71(6)&  &(2.5,-0.5)& -3.76 & 2.12 & 1.15 & 6.5(0.6)& 8.7(0.8)& 22(2)& 2.6 \\
\hline
\end{tabular}

$^1${$\star$ and $\Delta$ represent that this core is in Taurus and Cepheus region respectively.}
$^2${The uncertainty contributed by assuming a fixed abundance of $X$[\CCH] is not included.}

\end{table}
\end{landscape}

\end{document}